\documentclass[aps,pra,reprint,superscriptaddress]{revtex4-2}

\usepackage{graphicx}
\usepackage{latexsym}
\usepackage{dcolumn}
\usepackage[utf8]{inputenc}
\usepackage[T1]{fontenc}
\usepackage{tabularx}
\usepackage{hyperref}
\usepackage{color}
\usepackage{placeins}
\usepackage{float}
\usepackage{amsmath}
\usepackage{gensymb}

\begin{document}

\title{Computational characterization of novel nanostructured materials: A case study of NiCl$_2$}

\author{Elizaveta B. Kalika}
\affiliation{Moscow Institute of Physics and Technology, Institutskiy per. 9, Dolgoprudny, Moscow Region, 141700, Russia}
\affiliation{Research Institute for the Development of Scientific and Educational Potential of Youth, Aviatorov Str. 14/55, Moscow 119620, Russia}

\author{Alexey V. Verkhovtsev}
\affiliation{MBN Research Center, Altenh{\"o}ferallee 3, 60438 Frankfurt am Main, Germany}

\author{Mikhail M. Maslov}
\affiliation{National Research Nuclear University “MEPhI”, Kashirskoe Shosse 31, Moscow, 115409, Russia}
\affiliation{Research Institute for the Development of Scientific and Educational Potential of Youth, Aviatorov Str. 14/55, Moscow 119620, Russia}

\author{Konstantin P. Katin}
\affiliation{National Research Nuclear University “MEPhI”, Kashirskoe Shosse 31, Moscow, 115409, Russia}
\affiliation{Research Institute for the Development of Scientific and Educational Potential of Youth, Aviatorov Str. 14/55, Moscow 119620, Russia}

\author{Andrey V. Solov'yov}
\affiliation{MBN Research Center, Altenh{\"o}ferallee 3, 60438 Frankfurt am Main, Germany}

\begin{abstract}
A computational approach combining dispersion-corrected density functional theory (DFT) and classical molecular dynamics is employed to characterize the geometrical and thermo-mechanical properties of a recently proposed 2D transition metal dihalide NiCl$_2$. The characterization is performed using a classical interatomic force field whose parameters are determined and verified through the comparison with the results of DFT calculations. The developed force field is used to study the mechanical response, thermal stability, and melting of a NiCl$_2$ monolayer on the atomistic level of detail. The 2D NiCl$_2$ sheet is found to be thermally stable at temperatures below its melting point of $\sim$695~K. At higher temperatures, several subsequent structural transformations of NiCl$_2$ are observed, namely a transition into a porous 2D sheet and a 1D nanowire. The computational methodology presented through the case study of NiCl$_2$ can also be utilized to characterize other novel 2D materials, including recently synthesized NiO$_2$, NiS$_2$, and NiSe$_2$.
\end{abstract}

\maketitle

\section{Introduction}
\label{sec:Intro}

Nickel-based nanomaterials are considered promising for sensors, adsorbents, and catalysts. Recently synthesized Ni-based composites have proven their effectiveness as sensors for various organic and biological molecules \cite{deng2021, KARIMIMALEH2020123042, AHMAD2019139}.
Nickel nanoparticles possess catalytic activity for many reactions, including the synthesis of primary amines and hydrogen oxidation \cite{jagadeesh2019, OSHCHEPKOV2018447}.
Nickel-based two-dimensional (2D) materials can be used for the efficient generation of molecular oxygen \cite{Luan2018, wang}, electrocatalytic water splitting \cite{Liang2019}, and glucose oxidation \cite{Shu2018}. In addition, recently synthesized pentagonal 2D sheets of nickel diazenide (NiN$_2$) have a tunable direct band gap and may serve as a precursor for pentagonal 2D materials \cite{Bykov2021}.

Nickel-doped graphene materials are also widely used for many applications. In particular, they provide excellent sensing properties \cite{Phan2014, Ren2019, Deokar2020} and can effectively catalyze technologically important reactions, such as hydrogen production from water \cite{Zhang2017} and ethanol steam \cite{Chen2019},
oxygen reduction under alkaline conditions \cite{Toh2013}, and carbon monoxide reduction \cite{Xu2016}.
Another important application of Ni-doped graphene is hydrogen storage \cite{Zhou2016, Zhou2021, Safina2021}. Due to 3d electrons and the spillover effect, nickel can hold hydrogen molecules. At the same time, graphene is not the only 2D material that can be doped with nickel. Other 2D materials, such as silicene, germanene and MoS$_2$, have also been used as a substrate for nickel atoms and nanoparticles \cite{Manjanath2014, Li2018, Li2018a}.

The aforementioned properties of Ni-doped 2D materials are based on a combination of the mechanical characteristics of a 2D sheet with the adsorption and catalytic properties of nickel. A further development of this idea is that nickel can be not only an alloying impurity but also the basis for a 2D material. Pure nickel does not exist in the form of a 2D allotrope. However, an extensive computational search for layered crystals and related 2D materials \cite{Mounet2018} has recently predicted the existence of several 2D nickel-based materials of the composition NiX$_2$ (X = O, Cl, Br, I, S, and Se). Some of these materials have already been synthesized. Technologies developed for widespread 2D MoS$_2$ and MoSe$_2$ materials \cite{McCreary2014} and their analogs based on niobium \cite{Wang2017}, titanium \cite{Sherrell2018} and tungsten \cite{Cong2013} have been proven useful for synthesizing NiX$_2$ monolayers. Large-area NiO$_2$ monolayers separated by lanthanum atoms have been observed using scanning tunneling microscopy \cite{Ikeda2016}. Recent experiments have proven the feasibility of chemical vapor deposition synthesis of nanometer-sized layers of NiCl$_2$ \cite{Luo2020}, NiS$_2$ \cite{Dai2020}, and NiSe$_2$ \cite{Liu2018}. Moreover, NiCl$_2$ films up to four layers thick have been synthesized recently \cite{Luo2020}. In all these materials, nickel atoms are organized in a regular 2D lattice so that the materials possess enhanced adsorption and catalytic properties.

Currently, there is limited experimental information on the properties of such 2D materials since only a few laboratories have synthesized them in limited quantities.
Atomistic computer simulations can serve as a useful complementary approach to characterize the structure of such novel materials and explore their properties.

This paper presents the results of a computational characterization of a novel 2D material, NiCl$_2$, using a multiscale modeling approach. NiCl$_2$ has been chosen as an illustrative and experimentally relevant representative of the Ni-based 2D materials family. A combination of quantum and classical approaches into a unified multiscale methodology permits a comprehensive investigation of the structure, mechanical properties, and thermal stability of NiCl$_2$.

Density-functional theory (DFT) calculations provide reference data on the geometrical characteristics of the 2D NiCl$_2$ material. On this basis, a new classical interatomic potential is developed and benchmarked against the results of quantum-mechanical calculations. The validated potential is used for atomistic modeling of the mechanical and thermal stability of NiCl$_2$ using classical molecular dynamics (MD) employing the MBN Explorer \cite{Solovyov2012} and MBN Studio \cite{Sushko2019} software packages.

Through an illustrative case study of NiCl$_2$, we present a general methodology that can be utilized for the computational characterization of other novel materials, including the recently synthesized NiO$_2$ \cite{Ikeda2016}, NiS$_2$ \cite{Dai2020}, and NiSe$_2$ \cite{Liu2018}.
To the best of our knowledge, a 2D NiCl$_2$ has been studied computationally using DFT only in a few studies \cite{Lu_2019_ACSOmega.4.5714, Kistanov_2022_JPCL}, and there have been no MD-based studies of this material or other 2D nickel dihalides. The paper \cite{Lu_2019_ACSOmega.4.5714} studied the mechanical, electronic, and magnetic properties of monolayer and bilayer NiX$_2$ (X = Cl, Br, I) 2D structures. The recent study \cite{Kistanov_2022_JPCL}
investigated structural defects in NiCl$_2$ and similar materials and confirmed their stability in the environment using DFT calculations.

The paper is organized as follows. Section~\ref{sec:Methodology} describes the key aspects of theoretical methods utilized in DFT calculations and classical simulations. A particular focus is made on describing the procedure to determine the classical force field parameters for NiCl$_2$.
This follows with the discussion of the obtained results in Section~\ref{sec:Results}.
The accuracy of the developed force field is evaluated in Section~\ref{sec:Results_benchmark-FF} through the analysis of structural and energetic parameters of NiCl$_2$.
Section~\ref{sec:Results_mechanical_DFT} is devoted to the analysis of the mechanical properties of NiCl$_2$ by means of DFT and classical energy minimization calculations using the developed force field.
In Section~\ref{sec:Results_thermomech_MD}, the force field is utilized to study the thermal stability of NiCl$_2$ and evaluate its melting temperature using classical MD simulations.
Finally, in Section~\ref{sec:Conclusions}, we draw a conclusion from this work and give an outlook for further developments in this research direction.

\section{Computational Methodology}
\label{sec:Methodology}

\subsection{DFT calculations}
\label{sec:Methodology_DFT}

DFT calculations of the structural properties of a NiCl$_2$ sheet have been performed using the QUANTUM Espresso 6.5 software package \cite{Giannozzi2009, Giannozzi2017}. The plane-wave basis set for valence electron states, generalized gradient approximation (GGA) in the Perdew-Burke-Ernzerhof (PBE) functional form for the exchange-correlation energy \cite{Perdew1996}, and projector-augmented-wave (PAW) pseudopotentials \cite{Bloechl1994,Kresse1999} for core-electron interactions were used to perform the calculations. We have employed the kinetic energy cut-off for wave functions of 100~Ry (1360~eV) and kinetic energy cut-off for charge density and potential of 600~Ry (8160~eV) with checking the convergence of energy and charge to increase the simulation accuracy. In addition, the van der Waals interactions have been considered through the D3 Grimme (DFT-D3) dispersion corrections \cite{Grimme2010}. The DFT-D3 approach possesses improved accuracy due to using environment-dependent dispersion coefficients and the inclusion of a three-body component to the dispersion correction energy term.

\begin{figure*}[t!]
\centering
\includegraphics[width=0.9\textwidth]{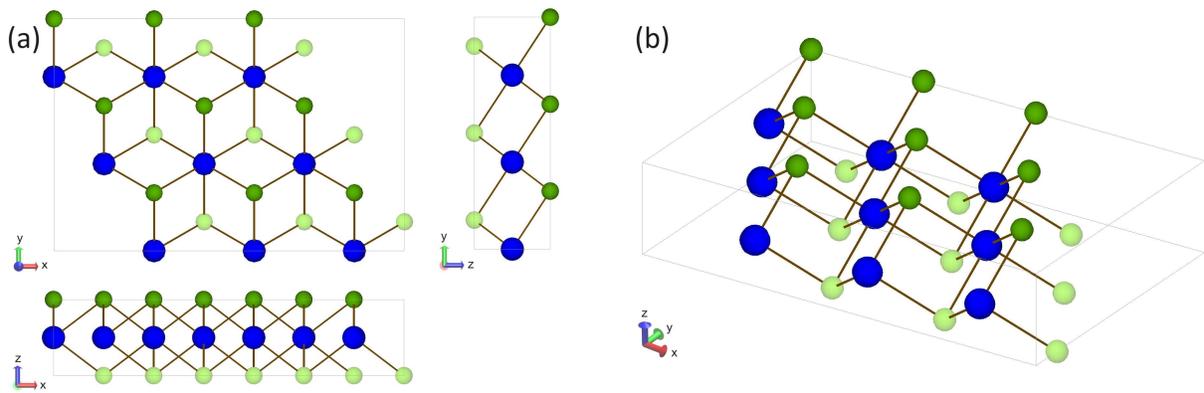}
\caption{Structure of a NiCl$_2$ supercell used in the DFT calculations. The supercell contains 9 nickel and 18 chlorine atoms, shown in blue and green (dark and light) colors, respectively. Chlorine atoms in the top and bottom atomic layers are shown by dark-green and light-green colors, respectively. Panel~(a) shows three orthographic projections of the considered supercell -- top and side views (top row) and a front view (bottom row). Panel~(b) shows a 3-dimensional projection of the supercell.}
\label{fig:NiCl2_supercell}
\end{figure*}

The distance between separate NiCl$_2$ sheets has been set equal to 30~\AA, which provides sufficient space separation to avoid nonphysical interactions. Thus, optimization of the lattice parameter along the axis perpendicular to the NiCl$_2$ sheet plane was unnecessary. The atomic equilibrium positions have been obtained by the total energy minimization of the supercell using the calculated forces and stress on the atoms. The convergence criterion for self-consistent calculations for ionic relaxations was set to $10^{-10}$~eV between two consecutive steps. The geometry optimization of the NiCl$_2$ sheet was carried out without symmetry constraints until the Hellman-Feynman forces acting on the atoms became smaller than $10^{-4}$ hartree/bohr. Such criteria ensure that the absolute value of stress is less than 0.01 kbar. The parameters of the supercell have also been optimized. The first Brillouin zone integrations have been performed using the Monkhorst-Pack $k$-point sampling scheme \cite{Monkhorst1976} with the $6 \times 6 \times 1$  mesh grid and the Methfessel-Paxton smearing \cite{Methfessel1989} with the smearing width of 0.02~Ry.

\begin{figure*}[t!]
\centering
\includegraphics[width=0.9\textwidth]{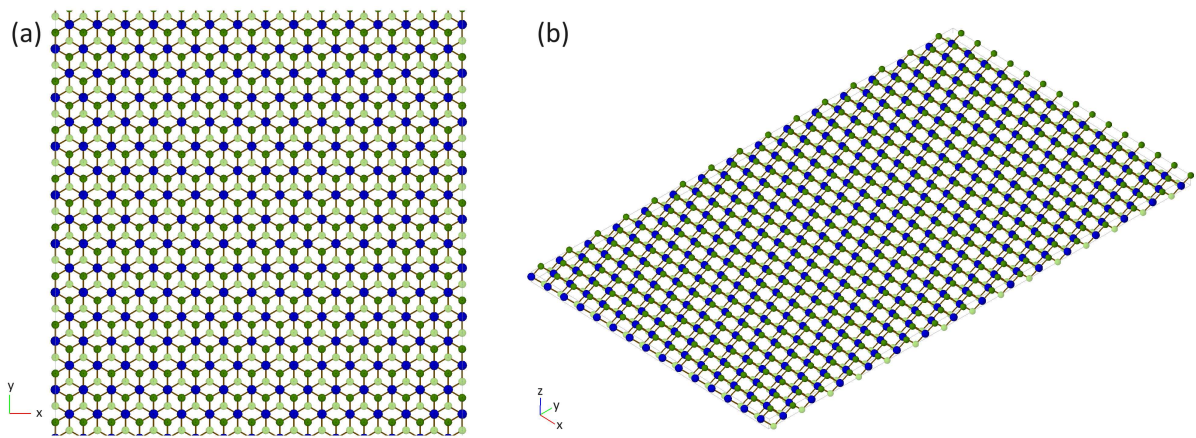}
\caption{Structure of a NiCl$_2$ sheet used in classical MD simulations. Panel~(a) shows the top view of a fragment of a NiCl$_2$ sheet, while panel~(b) shows a 3-dimensional projection of the entire system containing 1350 atoms. Ni and Cl atoms are shown in blue and green colors, respectively. Cl atoms in the top and bottom atomic layers are shown by dark-green and light-green colors, respectively.}
\label{fig:NiCl2_sheet_MD}
\end{figure*}

The NiCl$_2$ sheet is represented by a hexagonal Ni$_9$Cl$_{18}$ periodic cell containing $3 \times 3$ elementary NiCl$_2$ cells, see Fig.~\ref{fig:NiCl2_supercell}. The hexagonal symmetry corresponds to the results obtained earlier for this monolayer \cite{Mounet2018}. The lattice constant determined from the geometry optimization calculation is $a = 3a_0 = 10.328$~\AA, where $a_0$ is the parameter of the NiCl$_2$ primitive unit cell. The equilibrium Ni--Cl bond length obtained from the DFT calculations is equal to 2.38~\AA.

\subsection{Classical MD simulations}
\label{sec:Methodology_MBN}

Classical geometry optimization calculations and MD simulations have been performed by means of MBN Explorer \cite{Solovyov2012} -- a software package for multiscale modeling of complex molecular structures and dynamics. MBN Explorer permits the simulation of a wide range of Meso-Bio-Nano (MBN) systems \cite{MBNbook_Springer_2017, DySoN_book_Springer_2022}, including nanosystems, nanostructured materials, as well as composite and hybrid materials with sizes ranging from atomic to mesoscopic \cite{Verkhovtsev_2014_CNT, Sushko_2014_JPCA_FF, Sushko_2016_IDMD, deVera_2020_FEBID, Geng_2009_C60-TMB_JPCC, Moskovkin_2014}. The dedicated MBN Studio toolkit \cite{Sushko2019} has been utilized to create the systems, prepare all necessary input files, and analyze and visualize simulation outputs.

\subsection{Determination of the classical force field for NiCl$_2$}
\label{sec:Methodology_FF_param}

The first part of this study has been devoted to determining the parameters of a classical force field for NiCl$_2$.
The geometry of the NiCl$_2$ supercell obtained after DFT-based optimization (see Fig.~\ref{fig:NiCl2_supercell}) was taken as an initial geometry for the calculations using the classical force field.
The NiCl$_2$ supercell was translated in space along the translation vectors, and the generated crystal was cut to fill in a rectangular simulation box, which was replicated in space using periodic boundary conditions.
The resulting structure of a 2D NiCl$_2$ system is shown in Fig.~\ref{fig:NiCl2_sheet_MD}.
The simulation box used for the classical simulations has the size of $52.08~\textrm{\AA}~\times 90.30~\textrm{\AA}~\times 100.00~\textrm{\AA}$ and contains 1350 atoms.

\subsubsection{Determination of parameters through classical MD simulations}
\label{sec:FF_param_MD}

As a first educated guess, several MD simulations have been carried out with trial force field parameters fitted for other transitional metal halides NiF$_2$ \cite{nif2}, ZnCl$_2$ \cite{zncl2}, and FeCl$_3$ \cite{alcl3}.
The force field employed in these simulations has been constructed as a sum of the short-term exponential, long-term $r^{-6}$ power, and the Coulomb potentials.
Such force fields have been commonly utilized in MD simulations of metal halides and oxides, as well as various ionic crystals and other ionocovalent systems (e.g. glasses) \cite{Matsui_1991, Mattoni_2015_JPCC.119.17421, Morgan_2018, Pedone_2022}.
The NiCl$_2$ system was unstable with any considered trial parameters; the sheet lost its planarity and became strongly bent using the considered force field potentials.

Therefore, a new set of classical force field parameters for the 2D NiCl$_2$ system has been determined in this study as follows. The Nelder-Mead algorithm \cite{nelder-mead} has been utilized to optimize the force field parameters. The algorithm is simplex-based, meaning that in a $n$-dimensional space, it maintains a set of $n+1$ points called simplex. After calculating the value of the function at each point of the simplex, the algorithm extrapolates the function's behavior to find a new test point and replaces the test points with the worst result with a new one. This process continues until the result converges to the pre-defined tolerance value or reaches the maximum number of iterations.
The Nelder-Mead algorithm was implemented in Python via scipy.optimize.minimize library \cite{python}. This algorithm was used to fit the geometrical parameters of NiCl$_2$ calculated using MBN Explorer \cite{Solovyov2012} to the corresponding parameters obtained through the DFT-based structure optimization; results of this analysis are presented below in Sect.~\ref{sec:Results_benchmark-FF}.

NiCl$_2$ is an ionic crystal where polarization effects due to the ions of opposite charges may play a significant role. Several computational approaches for describing the ion-induced polarization interaction in various molecular systems within the classical MD framework have been discussed in the literature, including the addition of a $r^{-4}$ term to a $r^{-6}$ long-range potential \cite{Li_2015_JPCB.119.883, Turupcu_2020_JCTC.16.7184, Li_2017_ChemRev.117.1564} or considering a $r^{-4}$ term as the only contribution to the long-range interaction \cite{Mason_1972_JPB.5.169, Nelson_2001_JPD.34.3247}. In this study, we have used the latter approach for the sake of minimizing the number of force field parameters to be determined.
The resulting total interaction potential between atoms $i$ and $j$ of the system is given by the following expression:
\begin{equation}
U(r_{ij}) = A_{ij} \, e^{-\alpha_{ij} r_{ij}} - \frac{C_{ij}}{r_{ij}^4} + \frac{q_i q_j}{4 \pi \varepsilon_0 r_{ij}}
\label{eq:force_field_NiCl2}
\end{equation}
As shown below in Sections~\ref{sec:Results_benchmark-FF} and \ref{sec:Results_mechanical_DFT}, the force field given by Eq.~(\ref{eq:force_field_NiCl2}) provides a good overall agreement with the results of DFT calculations regarding the structural and mechanical properties of a 2D NiCl$_2$ material.

Six parameters of the force field, Eq.~(\ref{eq:force_field_NiCl2}), have been optimized using the Nelder-Mead algorithm: $A_{\rm Ni-Cl}$, $\alpha_{\rm Ni-Cl}$, $A_{\rm Cl-Cl}$, $\alpha_{\rm Cl-Cl}$, $C_{\rm Cl-Cl}$ and $q_{\rm Cl}$.
Partial charges on Ni and Cl atoms have been set to ensure the system is electrically neutral, so that $|q_{\rm Ni}| = 2|q_{\rm Cl}|$.
Following the approach adopted in several earlier studies \cite{Catlow_1982_Interionic, Gillan_1986_JPC.19.3391, BKS_SiO2_1990_PRL, nif2, Sundararaman_2018_JCP.148.194504} for a number of ionic crystals and ionocovalent systems, we have not considered the non-Coulombic contributions between the positively charged (in our case, nickel) ions, i.e.  $A_{\rm Ni-Ni} = 0$ and $C_{\rm Ni-Ni} = 0$. Following the considerations made in Ref.~\cite{Gillan_1986_JPC.19.3391}, the large partial atomic charge on the cations ensures that they are always well separated from each other, and their repulsive and van der Waals interactions can be neglected. The parameter $C_{\rm Ni-Cl}$ for the long-range interaction between the ions of different signs was also set equal to zero (see e.g. Refs. \cite{nif2, zncl2, alcl3}) on the grounds that unlike (i.e. Ni and Cl) nearest neighbours are close enough for the van der Waals interaction to be quenched \cite{Catlow_1982_Interionic, Gillan_1986_JPC.19.3391}.
A 7~\AA~cutoff was applied for the exponential and power terms of the potential. The Coulomb potential was calculated using the Ewald summation method \cite{Ewald_1921_AnnPhys.369.253, MBNExplorer_UserGuide} with the cutoff of 12~\AA. MD simulations with a simulation time of 30~ps and a time step of 1~fs, Langevin thermostat temperature of 300~K, and damping time of 0.1~ps were performed for each generated set of parameters.

The following geometrical parameters of NiCl$_2$ have been compared to the results of DFT calculations:
(i) Ni--Cl bond length ($d_{\rm Ni-Cl}$);
(ii) Ni--Ni bond length ($d_{\rm Ni-Ni}$);
(iii) in-plane Cl--Cl bond length ($d_{\rm Cl-Cl}$);
and (iv) the interplanar Cl--Ni--Cl angle ($\varphi_{\rm Cl-Ni-Cl}$), denoted hereafter as $\varphi$ for brevity.
In addition, the system's energy as a function of the Ni--Cl interplanar distance $l$ has also been analyzed. The distance between the Ni and Cl atomic planes was gradually varied, and single-point energy calculations were carried out using MBN Explorer for different values of $l$. The resulting dependence of the system's energy on the Ni--Cl interplanar distance was fitted by a quadratic function. The value of $l$ corresponding to the minimum system's energy was determined from classical force field calculations and compared to the results of DFT calculations.
Then, the standard deviation
\begin{equation}
\sigma
=
\sqrt{\sum_i{ \left( \frac{d_i-d_i^{\rm DFT}}{d_i^{\rm DFT}} \right)^2}
+
\left( \frac{\varphi - \varphi^{\rm DFT}}{\varphi^{\rm DFT}} \right)^2
+
\left( \frac{l-l^{\rm DFT}}{l^{\rm DFT}} \right)^2}
\label{eq1}
\end{equation}
was calculated and minimized using the Nelder-Mead algorithm \cite{nelder-mead}. Here, the summation has been carried out over all Ni--Ni, Ni--Cl and Cl--Cl covalent bonds of the system. Each round of force field parameter optimization consisted of 30 iterations, and the parameters resulting from each optimization round were used as the starting parameters for the following one. 400 subsequent optimization rounds have been performed to determine the parameters of the force field.
The force field parameters obtained via classical MD simulations performed at 300~K are listed in Table~\ref{table:FF_param-1} in Appendix.

\begin{table}[t!]
\caption{Parameters of the force field, Eq.~(\ref{eq:force_field_NiCl2}), derived in this study for the description of a 2D NiCl$_2$ material (see the main text for details).}
\begin{center}
\begin{tabular}{c|c|c|c||c|c}
\hline
	& $A_{ij}$~(eV) & $\alpha_{ij}$~(\AA$^{-1}$) & $C$~(eV \AA$^4$) &  & $q$, $|e|$  \\
\hline
\hline
 Ni--Cl  &  10715.83 &  5.034  &  $-$     & Ni & 1.34  \\
 Cl--Cl  &   1439.08 &  3.494  &  $0.779$ & Cl & $-0.67$ \\
 \hline
\end{tabular}
\end{center}
\label{table:FF_param}
\end{table}

\subsubsection{Determination of parameters through structure optimization calculations}

Due to relatively long simulation times, the MD-based procedure described in the previous section is inefficient for scanning a multidimensional parameter surface over a broad range of parameters. Therefore, the second round of force field parameter optimization was performed through a series of classical structure optimization calculations. The parameters generated at the previous step (see Sect.~\ref{sec:FF_param_MD} and Table~\ref{table:FF_param-1} in Appendix) were used as a starting point. Similarly to the previous step, the force field optimization procedure was performed using the Nelder-Mead algorithm \cite{nelder-mead}. Classical structure optimization calculations were performed using the velocity quenching algorithm for 5000 optimization steps with a time step of 0.1~fs.

A much shorter computational time of a structure optimization calculation than an MD simulation has allowed us to explore a broader range of force field parameter values.
However, it was observed that the optimization of all force field parameters, including partial charges, produces unrealistic non-physical values of partial charges (much larger than the formal charges of $+2$ and $-1$ on Ni and Cl atoms, respectively) due to lowering the total potential energy of the system with an increase of attractive electrostatic interaction between positively charged Ni atoms and negatively charged Cl atoms. In order to avoid this non-physical behavior, partial charges on Ni and Cl have been considered as fixed parameters (ensuring the total charge of the whole system is zero), and ten independent rounds of force field parameter optimization have been performed with constant values of atomic partial charges $q_{\rm Cl}$ in the range from $-0.859|e|$ to $-0.5|e|$. The force field parameters obtained after each round of structure optimization calculations were tested by analyzing the stability and planarity of the NiCl$_2$ system over a one-nanosecond-long MD simulation at 300~K. The final choice of the atomic partial charges was made to get the closest correspondence to the DFT-based results on the geometrical and mechanical properties of the 2D NiCl$_2$ system, as described below in Sections~\ref{sec:Results_benchmark-FF} and \ref{sec:Results_mechanical_DFT}.

It should be noted that the question of choosing atomic partial charges for the description of ionic crystals and other ionocovalent systems (such as silicate glasses) has been discussed in the literature, see e.g. Ref. \cite{Liu_2020_JCP.152.051101} and references therein.
MD simulations of SiO$_2$ systems showed that using formal charges (i.e., $+4$ for Si and $-2$ for O atoms) requires the use of additional angular 3-body energy terms to properly describe the system's geometry \cite{Woodcock_1976_JCP.65.1565}. Therefore, a commonly adopted approach for simulating ionocovalent systems is to use partial (rather than formal) atomic charges that are treated as an additional fitting parameter \cite{BKS_SiO2_1990_PRL, Kramer_1991_PRB.43.5068, Sundararaman_2018_JCP.148.194504} in order to reproduce the correct system's geometry without introducing additional energy terms to the empirical force field. This approach has been adopted in the present study to determine the partial charges on nickel and chlorine atoms.

The final values of the force field parameters and atomic partial charges for NiCl$_2$, derived from the comparison of NiCl$_2$ geometrical parameters to the DFT results, are listed in Table~\ref{table:FF_param}. These values have been employed in all the simulations described below in Section~\ref{sec:Results}.

\section{Results and Discussion}
\label{sec:Results}

\subsection{Evaluation of the accuracy of the constructed force field}
\label{sec:Results_benchmark-FF}

NiCl$_2$ is a novel 2D material for which little information has been obtained experimentally and computationally. Hence, it is important to evaluate the accuracy of the developed classical force field in order to make computational predictions of the structural and thermo-mechanical properties of NiCl$_2$.

Table~\ref{DFT_comparison} lists the values of Ni--Cl, Ni--Ni and Cl--Cl bond lengths ($d_{{\rm Ni-Cl}}$, $d_{{\rm Ni-Ni}}$ and $d_{{\rm Cl-Cl}}$) as well as the Cl--Cl interplanar distance ($l_{{\rm Cl-Cl}}$) and the Cl--Ni--Cl angle ($\varphi_{{\rm Cl-Ni-Cl}}$) obtained in this study by means of classical force field optimization (column labeled `FF') and DFT optimization calculations using the PBE exchange-correlation functional.
The present results are compared with several recent DFT calculations of a free 2D NiCl$_2$ monolayer \cite{Kistanov_2022_JPCL, Lu_2019_ACSOmega.4.5714} and that encapsulated inside a metal-organic framework \cite{Alnemrat_2023}.

\begin{table}[h!]
	\caption{Comparison of geometrical parameters of a NiCl$_2$ 2D sheet after geometry optimization using the classical force field, Eq.~(\ref{eq:force_field_NiCl2}) (column labeled ``FF''), and DFT using the PBE exchange-correlation functional. The results obtained in this study are also compared with the results of recent DFT calculations \cite{Kistanov_2022_JPCL, Lu_2019_ACSOmega.4.5714, Alnemrat_2023}.}
\centering
	\begin{tabular}{c|c|c|c|c|c}
		\hline
	Optimization method	& FF   & \multicolumn{4}{c}{DFT}    \\ \hline
		&  \multicolumn{2}{c|}{this work} &  Ref.~\cite{Kistanov_2022_JPCL} & Ref.~\cite{Lu_2019_ACSOmega.4.5714} & Ref.~\cite{Alnemrat_2023}    \\ \hline	
		$d_{{\rm Ni-Cl}}$, \AA & 2.24 &  2.38  &  2.38 &  2.42 & 2.42 \\ \hline
		$d_{{\rm Ni-Ni}}$, \AA &  3.57  & 3.44 &  &  & 3.56  \\ \hline
  	$d_{{\rm Cl-Cl}}$ (in-plane), \AA &  3.47  & 3.44 &  &   & \\ \hline
		$l_{{\rm Cl-Cl}}$ (interplanar), \AA &  2.68  & 3.28 &  &  2.65 & \\ \hline
		$\varphi_{{\rm Cl-Ni-Cl}}$, deg & 73.5 & 87.2 &  & & \\ \hline
	\end{tabular}
	\label{DFT_comparison}
\end{table}

The results listed in Table~\ref{DFT_comparison} indicate a good overall agreement (with the relative discrepancy of $\sim$5\%) between the outcomes of classical structure optimization calculations and the DFT calculations performed in this study and reported in the literature \cite{Kistanov_2022_JPCL, Lu_2019_ACSOmega.4.5714, Alnemrat_2023}. The interlayer vertical distance between two chlorine planes, $l_{{\rm Cl-Cl}}$, obtained from the classical calculations is about 20\% shorter than the values obtained by DFT, while being in good agreement with the results of the recent DFT-based study \cite{Lu_2019_ACSOmega.4.5714}. A smaller Cl--Cl interplanar distance results in smaller values of the angles $\varphi_{{\rm Cl-Ni-Cl}}$ compared to the outcomes of DFT calculations.

Overall, the classical force field. Eq.~(\ref{eq:force_field_NiCl2}), with the parameters listed in Table~\ref{table:FF_param} describes with reasonable accuracy the geometrical parameters of a 2D NiCl$_2$ sheet. The range of discrepancies between the results of classical force field-based and DFT-based geometry optimization calculations is similar to that obtained in earlier studies of transition-metal clusters and bulk materials \cite{Verkhovtsev_2013_TiNi_clusters, Sushko_2014_JPCA_FF}.

\subsection{Mechanical properties of NiCl$_2$}
\label{sec:Results_mechanical_DFT}

In this study, the mechanical properties of a NiCl$_2$ sheet have been analyzed by both DFT and classical force field calculations.
DFT-based calculations have used, as an input, the optimized NiCl$_2$ supercell described above in Section~\ref{sec:Methodology_DFT}.

The uniform stretching of the NiCl$_2$ sheet has been simulated by simultaneously increasing the lattice parameters of the supercell along the $x-$ and $y-$directions (parallel to the NiCl$_2$ plane) and further optimizing atomic positions in the supercell. Classical geometry optimization calculations were performed by means of MBN Explorer \cite{Solovyov2012} using the velocity quenching algorithm for 20,000 optimization steps with a time step of 1~fs.

During the biaxial stretching, the lattice constant $a$ increases as
\begin{equation}
a = a_0(1 + \varepsilon) \ ,
\label{eq:stretching_deformation}
\end{equation}
where $a_0$ is the equilibrium lattice constant and $\varepsilon$ is the relative deformation (strain).
The deformation energy $\Delta E$ has been calculated as follows:
\begin{equation}
    \Delta E = E - E_0 \ ,
\end{equation}
where $E$ is the total potential energy of the deformed system and $E_0$ is the potential energy of the unperturbed system.
As the simulation boxes considered in the DFT-based and classical force field optimization calculations contain different numbers of atoms, the calculated deformation energy was divided by the number of NiCl$_2$ units in the respective simulation box.

\begin{figure*}[t!]
\centering
\includegraphics[width=0.75\textwidth]{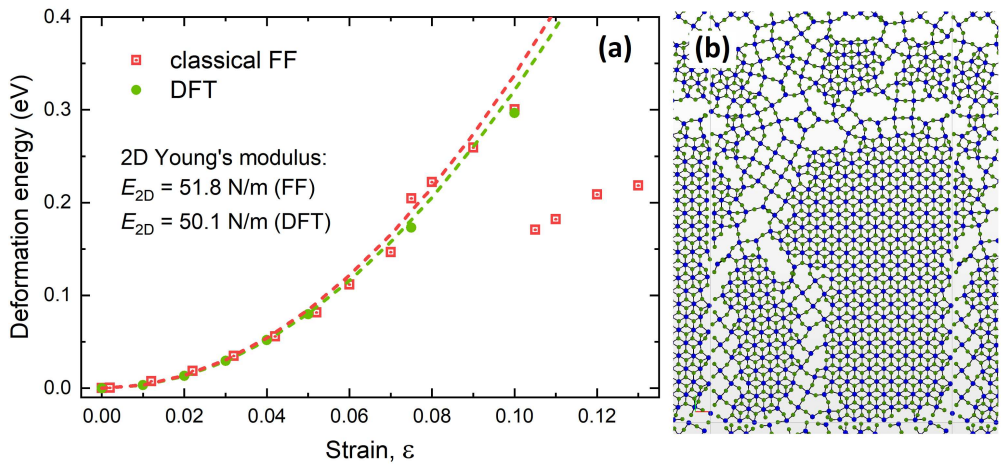}
\caption{Panel~(a): Deformation energy $\Delta E$ of a biaxially stretched NiCl$_2$ sheet as a function of relative deformation $\varepsilon$. Classical geometry optimization calculations (red symbols) have been performed using the force field, Eq.~(\ref{eq:force_field_NiCl2}), and the parameters listed in Table \ref{table:FF_param}. Panel~(b): Structure of NiCl$_2$ at strain $\varepsilon = 0.11$, which corresponds to the mechanical stability threshold for this material as predicted by the classical force field calculations. Ni and Cl atoms are shown in blue and green colors, respectively.}
\label{fig:stretching_energy}
\end{figure*}

Figure~\ref{fig:stretching_energy}(a) shows the dependence of the NiCl$_2$ sheet deformation energy on strain $\varepsilon$. In the region of small deformations (for strain values $\varepsilon \lesssim 0.05$), the dependence of deformation energy on $\varepsilon$ follows a quadratic dependence, indicative of the elastic deformation.
At larger deformations, the calculated dependence starts to deviate from the quadratic dependence. The calculations employing the classical force field, Eq.~(\ref{eq:force_field_NiCl2}), predict the deformation threshold at $\varepsilon \sim 0.11$, when the 2D sheet breaks into  smaller ``islands'' (see Fig.~\ref{fig:stretching_energy}(b)), which causes a sharp drop in the potential energy of the system and, hence, in its deformation energy.

The mechanical properties of 2D materials are commonly characterized via the 2D Young's modulus $E_{\rm 2D}$, which does not depend on the sheet thickness in contrast to the ``conventional'' 3D Young's modulus \cite{Kochaev2017-rm, Liu_2015_JMaterRes.31.832, Zhang_2022_JPCC.126.1094}.
The 2D Young's modulus is calculated as
\begin{equation}
E_{\rm 2D} = \frac{1}{2S_0} \, (\Delta E)'' \ ,
\end{equation}
where $S_0$ is the surface area of the relaxed 2D system and $(\Delta E)''$ is the second derivative of the deformation energy with respect to biaxial deformation $\varepsilon$.
A quadratic fit to the dependence of system's total deformation energy on $\varepsilon$, calculated by means of DFT, yields the value $E_{\rm 2D} = 50.1$~N/m. The classical force field, Eq.~(\ref{eq:force_field_NiCl2}), predicts a close value of 51.8~N/m. These values are in agreement with the value $E_{\rm 2D} = 54$~N/m, obtained for NiCl$_2$ using DFT calculations in Ref.~\cite{Lu_2019_ACSOmega.4.5714}. Similar values of the 2D Young's modulus were reported in the cited study \cite{Lu_2019_ACSOmega.4.5714} for related Ni-based 2D materials, NiBr$_2$ (50 N/m) and NiI$_2$ (45 N/m).
These results indicate that 2D NiCl$_2$ and other nickel dihalides are less rigid materials compared to other well-studied 2D sheets like graphene or MoS$_2$, which have the 2D Young's modulus of $\sim$340~N/m and 170~N/m, respectively \cite{Lee_2008_Science.321.385, Liu2014-ub}.

To estimate the ``conventional'' 3D Young's modulus, the obtained value $E_{\rm 2D}$ should be divided by the effective sheet thickness $h$.
According to the results obtained using the classical force field, Eq.~(\ref{eq:force_field_NiCl2}), the NiCl$_2$ sheet thickness (that is, the distance $l_{{\rm Cl-Cl}}$ between the chlorine planes) is equal to 2.68~\AA. The interlayer distance between the adjacent layers of NiCl$_2$ equals 3.08~\AA~\cite{Lu_2019_ACSOmega.4.5714} so that the effective sheet thickness can be estimated as $h = 5.76$~\AA. Using this value, one obtains the 3D Young's modulus $E_{\rm 3D} = E_{\rm 2D}/h \approx 90$~GPa.

\subsection{Melting temperature and thermal stability of NiCl$_2$}
\label{sec:Results_thermomech_MD}

\begin{figure}[t!]
\centering
\includegraphics[width=0.45\textwidth]{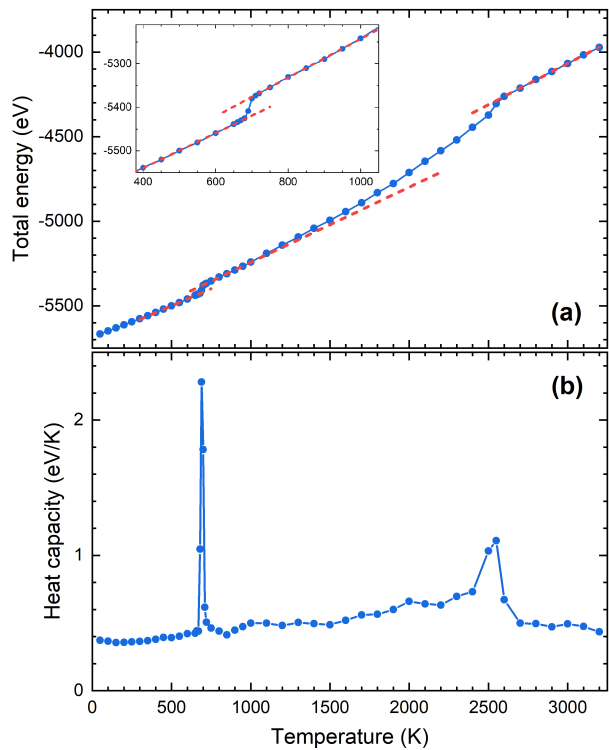}
\caption{Caloric curve for a 2D NiCl$_2$ material (panel~(a)) and the corresponding heat capacity $C_{\textrm v}$ as a function of temperature (panel~(b)) obtained using classical MD simulations. The peak in the $C_{\textrm v}(T)$ dependence at $T \approx 695$~K indicates the melting temperature of NiCl$_2$. A broader peak centered at $T \sim 2550$~K indicates a sublimation into the gas of NiCl$_2$ molecules and small atomic and molecular fragments (see the main text for details).}
\label{fig:caloric_curve}
\end{figure}

The interatomic force field described in Sect.~\ref{sec:Methodology_FF_param} has been utilized to study the thermal properties of a 2D NiCl$_2$ sheet and evaluate its melting temperature. This has been done by conducting a series of constant temperature MD simulations as follows.
The optimized structure of NiCl$_2$ was heated gradually by 50 degrees (in the temperature range from 0 to 650~K) or by 100 degrees (in the range from 800 to 3200~K) over 200~ps and then equilibrated over 1~ns at each temperature.
In order to avoid any artifacts related to the incomplete thermalization of the system, the first 400~ps of each trajectory simulated at a given temperature were excluded from the analysis. The average value of system's total energy at a given temperature was determined over the last 600-ps part of each trajectory. A smaller temperature increment of 10~K has been considered in the temperature range from 650 to 800~K where the melting phase transition occurs (see below). In this case, 10 ns-long simulations were performed at each temperature, and the system's total energy was analyzed over the last 6~ns of the simulated trajectories. The resulting caloric curve, that is the dependence of the time-averaged total energy of the system on temperature, $\langle E_{\rm tot}\rangle(T)$, is shown in Fig.~\ref{fig:caloric_curve}(a) by symbols. Figure~\ref{fig:caloric_curve}(b) shows the temperature dependence of heat capacity at constant volume $C_{\textrm v}$, defined as a derivative of the internal energy of the system over temperature:
\begin{equation}
C_{\textrm v} = \left(  \frac{\partial E}{\partial T} \right)_{\textrm v} \ .
\label{eq:heat-capacity}
\end{equation}
The melting process in a macroscopic system occurs at a specific temperature and reveals itself as a first-order phase transition via a spike in the heat capacity of the system at the transition temperature.

At the initial phase of NiCl$_2$ heating, the system's total energy grows linearly with temperature; see the inset in Fig.~\ref{fig:caloric_curve}(a). At temperatures above $\sim$600~K the crystalline structure of the NiCl$_2$ sheet is strongly deformed, but the system maintains its integrity.
A jump in the system's total energy has been observed at temperatures between 680 and 700~K.
The calculated heat capacity curve has a sharp maximum at $T \approx 695$~K, indicating the melting temperature of NiCl$_2$.
This value is lower than the melting temperature for bulk nickel, $T_{\rm m}({\rm Ni}) = 1728$~K, and of bulk NiCl$_2$, $T_{\rm m}({\rm NiCl_2^{bulk}}) = 1274$~K \cite{CRC_Handbook_Chemistry}.
At the same time, the evaluated melting temperature of 2D NiCl$_2$ is within the temperature range ($T \sim 500-1000$~K) where other 2D metal halides, such as NaCl, become thermally unstable and lose their structural integrity, as determined computationally from \textit{ab initio} MD simulations \cite{BingchengLuo_PNAS_2019}.

At temperatures above the melting point, the system's total energy $\langle E_{\rm tot}\rangle$ continues to grow linearly with $T$ (see the inset in Fig.~\ref{fig:caloric_curve}(a)) until the onset of another structural transformation emerges at $T \sim 1500$~K. In contrast to the melting phase transition at $T_{\rm m} \approx 695$~K, this high-temperature transformation takes place over a broad temperature range from approximately 1500 to 2700~K, as indicated by the deviation of the $\langle E_{\rm tot}\rangle(T)$ dependence from a linear function and a peak in the $C_{\textrm v}(T)$ dependence in the temperature range $\sim 2400-2700$~K. This transition corresponds to the evaporation of NiCl$_2$ fragments from a 1D NiCl$_2$ nanowire (see the discussion below).

\begin{figure*}[t!]
\centering
\includegraphics[width=0.8\textwidth]{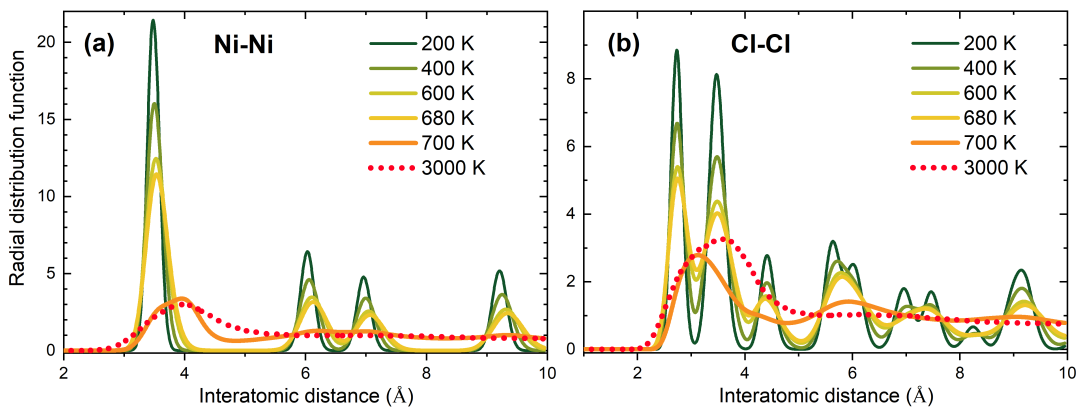}
\caption{Radial distribution function for (a) nickel and (b) chlorine atoms of NiCl$_2$ at different temperatures of the system. A rapid change of the RDFs between 680 and 700~K indicates the melting phase transition. }
\label{fig:RDF_Ni_Cl}
\end{figure*}

The structural transformations corresponding to the peaks in the heat capacity curve can be visualized by analyzing the radial distribution function (RDF) of atoms in the system. The RDF characterizes the number of atoms located at a certain radial distance $r$ from a reference atom.
The RDF for Ni--Ni and Cl--Cl atomic pairs, evaluated at different temperatures of the system, are plotted in Figures~\ref{fig:RDF_Ni_Cl}(a) and \ref{fig:RDF_Ni_Cl}(b), respectively. A rapid change of the RDFs between 680 and 700~K indicates the melting phase transition.

\begin{figure}[t!]
\centering
\includegraphics[width=0.48\textwidth]{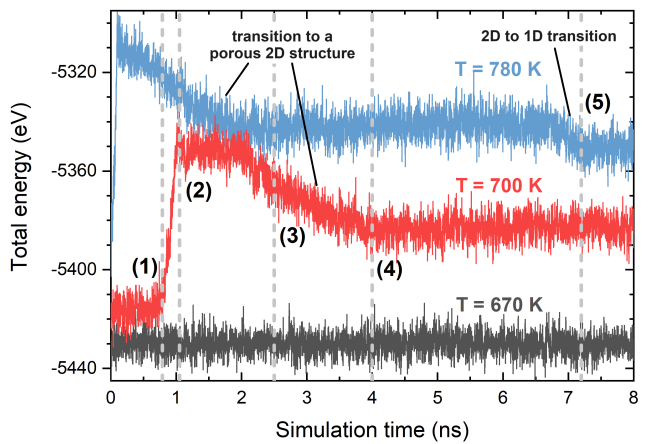}
\caption{Variation of the system's total energy, $E_{\rm tot}$, as a function of simulation time $t$. Colored curves show the $E_{\rm tot}(t)$ dependencies for different temperatures, namely $T = 670$~K (below the phase transition temperature), $T = 700$~K (in the phase transition region), and $T = 780$~K (above the phase transition temperature). Labels (1) to (4) correspond to different time instants (dashed lines) for the curve at $T = 700$~K, while label~(5) indicates the time instant for the curve at $T = 780$~K. These instants (1)--(5) correspond to the system's snapshots shown in panels~(c)--(g) of Fig.~\ref{fig:NiCl2_melting_snapshots}. }
\label{fig:Etot_time}
\end{figure}

It is of particular interest to explore in more detail the phase transition from the crystalline NiCl$_2$ to its molten state. Figure~\ref{fig:Etot_time} shows the variation of the system's total energy, $E_{\rm tot}$, as a function of simulation time $t$; the first 8~ns of the simulations (out of 10~ns) are shown for better clarity. Colored curves show the $E_{\rm tot}(t)$ dependencies for three different temperatures, namely $T = 670$~K (below the phase transition temperature), $T = 700$~K (in the phase transition region), and $T = 780$~K (above the phase transition temperature).

\begin{figure*}[t!]
\centering
\includegraphics[width=1.0\textwidth]{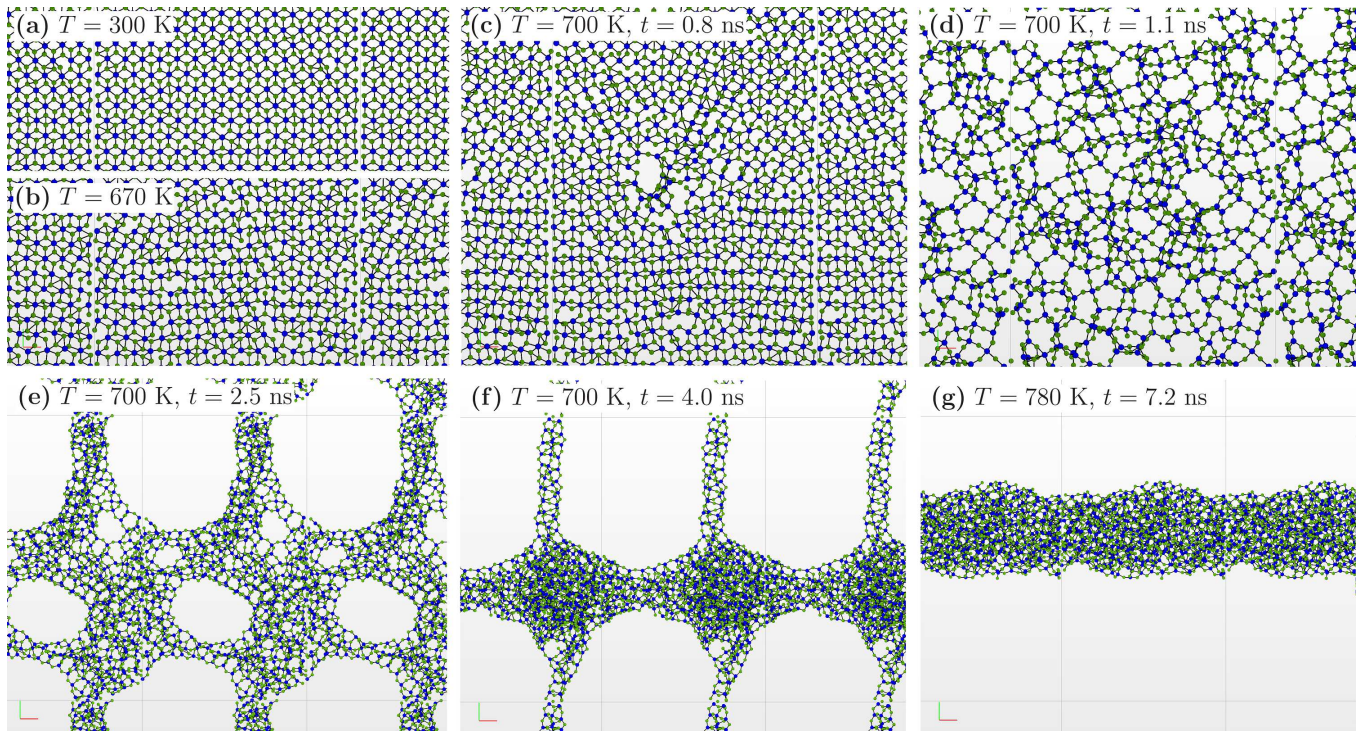}
\caption{Snapshots of the NiCl$_2$ system at different temperatures $T$ as indicated. Nickel and chlorine atoms are shown in blue and green colors, respectively.
Panels (c) to (g) illustrate the system's structure at different time instants corresponding to labels (1)--(5) in Fig.~\ref{fig:Etot_time}.}
\label{fig:NiCl2_melting_snapshots}	
\end{figure*}

\begin{figure*}[t!]
\centering
\includegraphics[width=1.0\textwidth]{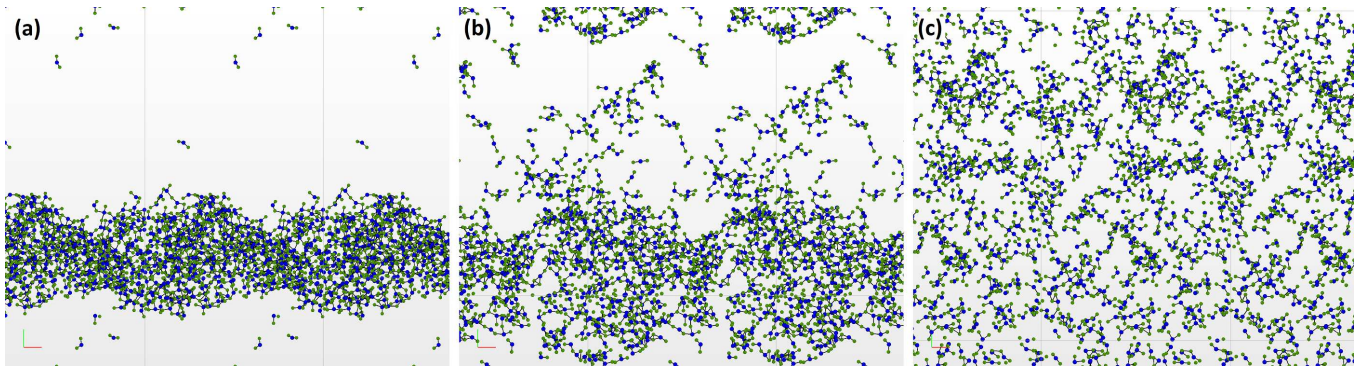}
\caption{Snapshots of the NiCl$_2$ system at temperatures $T = 1800$~K (a), 2400~K (b) and 3000~K (c) at the end of 1-nanosecond-long simulations. As the temperature increases, more NiCl$_2$ units are evaporated from the 1D NiCl$_2$ nanowire [see Fig.~\ref{fig:NiCl2_melting_snapshots}(g)], leading to its sublimation into a gas of NiCl$_2$ molecules and smaller fragments such as NiCl and isolated Ni and Cl atoms at $T \sim 2550$~K. Nickel and chlorine atoms are shown in blue and green colors, respectively.}
\label{fig:NiCl2_multifragmentation}	
\end{figure*}

At $T = 670$~K (black curve in Fig.~\ref{fig:Etot_time}), the total energy of NiCl$_2$ fluctuates around the average value of about $-5430$~eV, which indicates that the system is thermally stable at this temperature over the considered simulation time. At $T = 700$~K (red curve in Fig.~\ref{fig:Etot_time}), the system's energy stays nearly constant within the first $\sim$0.7~ns of the simulated trajectory, followed by a rapid rise of $E_{\rm tot}(t)$ within the next $\sim$0.4~ns. This jump in the total energy is attributed to the formation of defects (holes) in the NiCl$_2$ sheet, as shown in Fig.~\ref{fig:NiCl2_melting_snapshots}(c). After that, the system's total energy decreases as the system relaxes into a porous 2D structure, see Fig.~\ref{fig:NiCl2_melting_snapshots}(d,e). The structure shown in Fig.~\ref{fig:NiCl2_melting_snapshots}(e) is metastable and evolves into a different 2D structure, where denser regions are connected by thin NiCl$_2$ links, see Fig.~\ref{fig:NiCl2_melting_snapshots}(f).
It should be noted that the results presented in Figures~\ref{fig:Etot_time} and \ref{fig:NiCl2_melting_snapshots} have been obtained for the NiCl$_2$ sheet of a particular size, containing 1350 atoms.
A detailed analysis of the system's size on its thermal properties and the observed structural transformations is an interesting topic which, however, goes beyond the scope of the present study.

The structure shown in Fig.~\ref{fig:NiCl2_melting_snapshots}(f) is stable at $T = 700$~K over a 10~ns-long simulation. However, the ``dissolution'' of the thin NiCl$_2$ links seen in Fig.~\ref{fig:NiCl2_melting_snapshots}(f) might take place on a longer time scale, leading to the formation of a 1D structure. Analyzing the dynamics of the NiCl$_2$ system at higher temperatures enables to observe a 2D-to-1D structural transformation within the considered simulation times.
Indeed, at $T = 780$~K, the NiCl$_2$ system transforms into a 1D nanowire (see Fig.~\ref{fig:NiCl2_melting_snapshots}(g)), which is thermally stable at temperatures below $\sim$1500~K within the considered simulation times. This transformation is characterized by a decrease of the system's total energy, see the blue curve in Fig.~\ref{fig:Etot_time}.

\begin{sloppypar}
At temperatures of $\sim$1500~K, NiCl$_2$ monomers begin to evaporate from the NiCl$_2$ nanowire (see Fig.~\ref{fig:NiCl2_multifragmentation}(a)), which causes a deviation from the linear dependence in the caloric curve shown in Fig.~\ref{fig:caloric_curve}(a). As the temperature increases, more NiCl$_2$ units detach from the nanowire, resulting in the sublimation of the nanowire into a gas of NiCl$_2$ molecules and smaller fragments such as NiCl dimers and isolated Ni and Cl atoms, see Fig.~\ref{fig:NiCl2_multifragmentation}(b,c). This phenomenon is characterized by the change of the $\langle E_{\rm tot}\rangle(T)$ slope shown in Fig.~\ref{fig:caloric_curve}(a) and a peak in the heat capacity curve centered at $T \approx 2550$~K.
\end{sloppypar}

\section{Conclusions}
\label{sec:Conclusions}

Two-dimensional (2D) materials possess unique technologically-relevant properties, which are often quite different from the properties of the corresponding bulk counterparts. Monolayers of layered crystals are particularly interesting because they can be synthesized with a well-known `top-down' approach. Extensive experimental characterization of such materials is expensive and time-consuming, while computational modeling may provide valuable support for experimental research.

This study has reported results of a computer simulation of the recently proposed novel 2D material NiCl$_2$ by means of DFT calculations and classical MD simulations. In order to characterize the geometrical and thermo-mechanical properties of the 2D NiCl$_2$ system, a classical interatomic force field was developed and verified through the comparison with the results of DFT calculations.

The combination of a DFT approach and a classical approach based on an interatomic force field provides insight into the structural and thermo-mechanical properties of the 2D NiCl$_2$ material. It has been found that the NiCl$_2$ sheet is mechanically stable upon biaxial stretching at the relative deformation up to $11\%$. 2D Young's modulus for NiCl$_2$ obtained by DFT calculations and calculations employing the classical force field is $\sim50$~N/m, in agreement with the results of recent DFT calculations \cite{Lu_2019_ACSOmega.4.5714}. The Young's modulus for a 2D NiCl$_2$ system is about $5-10$ times lower than Young's modulus of other commonly studied 2D materials such as graphene, MoS$_2$ and WS$_2$. The performed MD simulations indicate that the NiCl$_2$ sheet is thermally stable up to the melting point temperature of $\sim 695$~K.
At temperatures above the melting point, structural degradation of NiCl$_2$ has been observed, which involves several subsequent structural transformations into a porous 2D sheet, a 1D nanowire, and a sublimation into a gas of NiCl$_2$ monomers and one- and two-atom fragments.

The classical force field developed in this study for simulating the 2D NiCl$_2$ material describes reasonably accurately the geometrical parameters of NiCl$_2$, determined on the basis of DFT calculations.
The force field parameters for NiCl$_2$ presented in this study might be useful for studying other characteristics of this material, such as thermal conductivity, elastic constants, or phonon dispersion curves.
Finally, the methodology presented through an illustrative case study of NiCl$_2$ can also be utilized for the computational characterization of other novel 2D materials, including recently synthesized NiO$_2$, NiS$_2$ and NiSe$_2$ materials.

\section*{Appendix}

Table~\ref{table:FF_param-1} lists trial parameters of the force field for a 2D NiCl$_2$ material, Eq.~(\ref{eq:force_field_NiCl2}), derived in this study through a series of constant-temperature MD simulations (see Sect.~\ref{sec:FF_param_MD} for details).

\begin{table}[h!]
\caption{Trial parameters of the force field for a 2D NiCl$_2$ material, Eq.~(\ref{eq:force_field_NiCl2}), derived in this study through a series of constant-temperature MD simulations.}
\begin{center}
\begin{tabular}{c|c|c|c||c|c}
\hline
	& $A_{ij}$~(eV) & $\alpha_{ij}$~(\AA$^{-1}$) & $C$~(eV \AA$^4$) &  & $q$, $|e|$  \\
\hline
 Ni--Cl  &  10804.52 &  4.73  &  $-$     & Ni & 1.718  \\
 Cl--Cl  &   1465.99 &  3.55  &  $0.805$ & Cl & $-0.859$ \\
 \hline
\end{tabular}
\end{center}
\label{table:FF_param-1}
\end{table}

\section*{Acknowledgements}

\begin{sloppypar}
The authors acknowledge the support of this work by the Deutsche Forschungsgemeinschaft (DFG), project no.~452691275.
The possibility of performing computer simulations at the Goethe-HLR cluster of the Frankfurt Center for Scientific Computing (CSC) is also gratefully acknowledged.
\end{sloppypar}

\bibliographystyle{apsrev4-2}
 \bibliography{NiCl2_references}

\begin{thebibliography}{84}%
\makeatletter
\providecommand \@ifxundefined [1]{%
 \@ifx{#1\undefined}
}%
\providecommand \@ifnum [1]{%
 \ifnum #1\expandafter \@firstoftwo
 \else \expandafter \@secondoftwo
 \fi
}%
\providecommand \@ifx [1]{%
 \ifx #1\expandafter \@firstoftwo
 \else \expandafter \@secondoftwo
 \fi
}%
\providecommand \natexlab [1]{#1}%
\providecommand \enquote  [1]{``#1''}%
\providecommand \bibnamefont  [1]{#1}%
\providecommand \bibfnamefont [1]{#1}%
\providecommand \citenamefont [1]{#1}%
\providecommand \href@noop [0]{\@secondoftwo}%
\providecommand \href [0]{\begingroup \@sanitize@url \@href}%
\providecommand \@href[1]{\@@startlink{#1}\@@href}%
\providecommand \@@href[1]{\endgroup#1\@@endlink}%
\providecommand \@sanitize@url [0]{\catcode `\\12\catcode `\$12\catcode
  `\&12\catcode `\#12\catcode `\^12\catcode `\_12\catcode `\%12\relax}%
\providecommand \@@startlink[1]{}%
\providecommand \@@endlink[0]{}%
\providecommand \url  [0]{\begingroup\@sanitize@url \@url }%
\providecommand \@url [1]{\endgroup\@href {#1}{\urlprefix }}%
\providecommand \urlprefix  [0]{URL }%
\providecommand \Eprint [0]{\href }%
\providecommand \doibase [0]{https://doi.org/}%
\providecommand \selectlanguage [0]{\@gobble}%
\providecommand \bibinfo  [0]{\@secondoftwo}%
\providecommand \bibfield  [0]{\@secondoftwo}%
\providecommand \translation [1]{[#1]}%
\providecommand \BibitemOpen [0]{}%
\providecommand \bibitemStop [0]{}%
\providecommand \bibitemNoStop [0]{.\EOS\space}%
\providecommand \EOS [0]{\spacefactor3000\relax}%
\providecommand \BibitemShut  [1]{\csname bibitem#1\endcsname}%
\let\auto@bib@innerbib\@empty
\bibitem [{\citenamefont {Deng}\ \emph {et~al.}(2021)\citenamefont {Deng},
  \citenamefont {Nie}, \citenamefont {Wu}, \citenamefont {Tian}, \citenamefont
  {Li},\ and\ \citenamefont {He}}]{deng2021}%
  \BibitemOpen
  \bibfield  {author} {\bibinfo {author} {\bibfnamefont {P.}~\bibnamefont
  {Deng}}, \bibinfo {author} {\bibfnamefont {X.}~\bibnamefont {Nie}}, \bibinfo
  {author} {\bibfnamefont {Y.}~\bibnamefont {Wu}}, \bibinfo {author}
  {\bibfnamefont {Y.}~\bibnamefont {Tian}}, \bibinfo {author} {\bibfnamefont
  {J.}~\bibnamefont {Li}},\ and\ \bibinfo {author} {\bibfnamefont
  {Q.}~\bibnamefont {He}},\ }\href@noop {} {\bibfield  {journal} {\bibinfo
  {journal} {Microchem. J.}\ }\textbf {\bibinfo {volume} {160}},\ \bibinfo
  {pages} {105744} (\bibinfo {year} {2021})}\BibitemShut {NoStop}%
\bibitem [{\citenamefont {Karimi-Maleh}\ \emph {et~al.}(2020)\citenamefont
  {Karimi-Maleh}, \citenamefont {Cellat}, \citenamefont {Arıkan},
  \citenamefont {Savk}, \citenamefont {Karimi},\ and\ \citenamefont
  {Şen}}]{KARIMIMALEH2020123042}%
  \BibitemOpen
  \bibfield  {author} {\bibinfo {author} {\bibfnamefont {H.}~\bibnamefont
  {Karimi-Maleh}}, \bibinfo {author} {\bibfnamefont {K.}~\bibnamefont
  {Cellat}}, \bibinfo {author} {\bibfnamefont {K.}~\bibnamefont {Arıkan}},
  \bibinfo {author} {\bibfnamefont {A.}~\bibnamefont {Savk}}, \bibinfo {author}
  {\bibfnamefont {F.}~\bibnamefont {Karimi}},\ and\ \bibinfo {author}
  {\bibfnamefont {F.}~\bibnamefont {Şen}},\ }\href@noop {} {\bibfield
  {journal} {\bibinfo  {journal} {Mater. Chem. Phys.}\ }\textbf {\bibinfo
  {volume} {250}},\ \bibinfo {pages} {123042} (\bibinfo {year}
  {2020})}\BibitemShut {NoStop}%
\bibitem [{\citenamefont {Ahmad}\ \emph {et~al.}(2019)\citenamefont {Ahmad},
  \citenamefont {Bedük}, \citenamefont {Majhi},\ and\ \citenamefont
  {Salama}}]{AHMAD2019139}%
  \BibitemOpen
  \bibfield  {author} {\bibinfo {author} {\bibfnamefont {R.}~\bibnamefont
  {Ahmad}}, \bibinfo {author} {\bibfnamefont {T.}~\bibnamefont {Bedük}},
  \bibinfo {author} {\bibfnamefont {S.~M.}\ \bibnamefont {Majhi}},\ and\
  \bibinfo {author} {\bibfnamefont {K.~N.}\ \bibnamefont {Salama}},\
  }\href@noop {} {\bibfield  {journal} {\bibinfo  {journal} {Sens. Actuators B
  Chem.}\ }\textbf {\bibinfo {volume} {286}},\ \bibinfo {pages} {139} (\bibinfo
  {year} {2019})}\BibitemShut {NoStop}%
\bibitem [{\citenamefont {Jagadeesh}\ \emph {et~al.}(2019)\citenamefont
  {Jagadeesh}, \citenamefont {Murugesan},\ and\ \citenamefont
  {Beller}}]{jagadeesh2019}%
  \BibitemOpen
  \bibfield  {author} {\bibinfo {author} {\bibfnamefont {R.}~\bibnamefont
  {Jagadeesh}}, \bibinfo {author} {\bibfnamefont {K.}~\bibnamefont
  {Murugesan}},\ and\ \bibinfo {author} {\bibfnamefont {M.}~\bibnamefont
  {Beller}},\ }\href@noop {} {\bibfield  {journal} {\bibinfo  {journal} {Angew.
  Chem. Int. Ed.}\ }\textbf {\bibinfo {volume} {58}},\ \bibinfo {pages} {5064}
  (\bibinfo {year} {2019})}\BibitemShut {NoStop}%
\bibitem [{\citenamefont {Oshchepkov}\ \emph {et~al.}(2018)\citenamefont
  {Oshchepkov}, \citenamefont {Bonnefont}, \citenamefont {Pronkin},
  \citenamefont {Cherstiouk}, \citenamefont {Ulhaq-Bouillet}, \citenamefont
  {Papaefthimiou}, \citenamefont {Parmon},\ and\ \citenamefont
  {Savinova}}]{OSHCHEPKOV2018447}%
  \BibitemOpen
  \bibfield  {author} {\bibinfo {author} {\bibfnamefont {A.~G.}\ \bibnamefont
  {Oshchepkov}}, \bibinfo {author} {\bibfnamefont {A.}~\bibnamefont
  {Bonnefont}}, \bibinfo {author} {\bibfnamefont {S.~N.}\ \bibnamefont
  {Pronkin}}, \bibinfo {author} {\bibfnamefont {O.~V.}\ \bibnamefont
  {Cherstiouk}}, \bibinfo {author} {\bibfnamefont {C.}~\bibnamefont
  {Ulhaq-Bouillet}}, \bibinfo {author} {\bibfnamefont {V.}~\bibnamefont
  {Papaefthimiou}}, \bibinfo {author} {\bibfnamefont {V.~N.}\ \bibnamefont
  {Parmon}},\ and\ \bibinfo {author} {\bibfnamefont {E.~R.}\ \bibnamefont
  {Savinova}},\ }\href@noop {} {\bibfield  {journal} {\bibinfo  {journal} {J.
  Power Sources}\ }\textbf {\bibinfo {volume} {402}},\ \bibinfo {pages} {447}
  (\bibinfo {year} {2018})}\BibitemShut {NoStop}%
\bibitem [{\citenamefont {Luan}\ \emph {et~al.}(2018)\citenamefont {Luan},
  \citenamefont {Liu}, \citenamefont {Liu}, \citenamefont {Yu}, \citenamefont
  {Wang}, \citenamefont {Xiao}, \citenamefont {Qiao}, \citenamefont {Dai},\
  and\ \citenamefont {Zhang}}]{Luan2018}%
  \BibitemOpen
  \bibfield  {author} {\bibinfo {author} {\bibfnamefont {C.}~\bibnamefont
  {Luan}}, \bibinfo {author} {\bibfnamefont {G.}~\bibnamefont {Liu}}, \bibinfo
  {author} {\bibfnamefont {Y.}~\bibnamefont {Liu}}, \bibinfo {author}
  {\bibfnamefont {L.}~\bibnamefont {Yu}}, \bibinfo {author} {\bibfnamefont
  {Y.}~\bibnamefont {Wang}}, \bibinfo {author} {\bibfnamefont {Y.}~\bibnamefont
  {Xiao}}, \bibinfo {author} {\bibfnamefont {H.}~\bibnamefont {Qiao}}, \bibinfo
  {author} {\bibfnamefont {X.}~\bibnamefont {Dai}},\ and\ \bibinfo {author}
  {\bibfnamefont {X.}~\bibnamefont {Zhang}},\ }\href@noop {} {\bibfield
  {journal} {\bibinfo  {journal} {ACS Nano}\ }\textbf {\bibinfo {volume}
  {12}},\ \bibinfo {pages} {3875} (\bibinfo {year} {2018})}\BibitemShut
  {NoStop}%
\bibitem [{\citenamefont {Wang}\ \emph {et~al.}(2019)\citenamefont {Wang},
  \citenamefont {Zhou}, \citenamefont {Han}, \citenamefont {Xi}, \citenamefont
  {You}, \citenamefont {Xianfeng}, \citenamefont {Li}, \citenamefont {Zhou},
  \citenamefont {Song}, \citenamefont {Wang},\ and\ \citenamefont
  {Gao}}]{wang}%
  \BibitemOpen
  \bibfield  {author} {\bibinfo {author} {\bibfnamefont {Y.}~\bibnamefont
  {Wang}}, \bibinfo {author} {\bibfnamefont {Y.}~\bibnamefont {Zhou}}, \bibinfo
  {author} {\bibfnamefont {M.}~\bibnamefont {Han}}, \bibinfo {author}
  {\bibfnamefont {Y.}~\bibnamefont {Xi}}, \bibinfo {author} {\bibfnamefont
  {H.}~\bibnamefont {You}}, \bibinfo {author} {\bibfnamefont {H.}~\bibnamefont
  {Xianfeng}}, \bibinfo {author} {\bibfnamefont {Z.}~\bibnamefont {Li}},
  \bibinfo {author} {\bibfnamefont {J.}~\bibnamefont {Zhou}}, \bibinfo {author}
  {\bibfnamefont {D.}~\bibnamefont {Song}}, \bibinfo {author} {\bibfnamefont
  {D.}~\bibnamefont {Wang}},\ and\ \bibinfo {author} {\bibfnamefont
  {F.}~\bibnamefont {Gao}},\ }\href@noop {} {\bibfield  {journal} {\bibinfo
  {journal} {Small}\ }\textbf {\bibinfo {volume} {15}},\ \bibinfo {pages}
  {1805435} (\bibinfo {year} {2019})}\BibitemShut {NoStop}%
\bibitem [{\citenamefont {Liang}\ \emph {et~al.}(2019)\citenamefont {Liang},
  \citenamefont {Zhong}, \citenamefont {Du}, \citenamefont {Luo}, \citenamefont
  {Zhao}, \citenamefont {Zheng}, \citenamefont {Xu}, \citenamefont {Ma},
  \citenamefont {Liu}, \citenamefont {Li},\ and\ \citenamefont
  {Yan}}]{Liang2019}%
  \BibitemOpen
  \bibfield  {author} {\bibinfo {author} {\bibfnamefont {Q.}~\bibnamefont
  {Liang}}, \bibinfo {author} {\bibfnamefont {L.}~\bibnamefont {Zhong}},
  \bibinfo {author} {\bibfnamefont {C.}~\bibnamefont {Du}}, \bibinfo {author}
  {\bibfnamefont {Y.}~\bibnamefont {Luo}}, \bibinfo {author} {\bibfnamefont
  {J.}~\bibnamefont {Zhao}}, \bibinfo {author} {\bibfnamefont {Y.}~\bibnamefont
  {Zheng}}, \bibinfo {author} {\bibfnamefont {J.}~\bibnamefont {Xu}}, \bibinfo
  {author} {\bibfnamefont {J.}~\bibnamefont {Ma}}, \bibinfo {author}
  {\bibfnamefont {C.}~\bibnamefont {Liu}}, \bibinfo {author} {\bibfnamefont
  {S.}~\bibnamefont {Li}},\ and\ \bibinfo {author} {\bibfnamefont
  {Q.}~\bibnamefont {Yan}},\ }\href@noop {} {\bibfield  {journal} {\bibinfo
  {journal} {ACS Nano}\ }\textbf {\bibinfo {volume} {13}},\ \bibinfo {pages}
  {7975} (\bibinfo {year} {2019})}\BibitemShut {NoStop}%
\bibitem [{\citenamefont {Shu}\ \emph {et~al.}(2018)\citenamefont {Shu},
  \citenamefont {Li}, \citenamefont {Chen}, \citenamefont {Xu}, \citenamefont
  {Pang},\ and\ \citenamefont {Hu}}]{Shu2018}%
  \BibitemOpen
  \bibfield  {author} {\bibinfo {author} {\bibfnamefont {Y.}~\bibnamefont
  {Shu}}, \bibinfo {author} {\bibfnamefont {B.}~\bibnamefont {Li}}, \bibinfo
  {author} {\bibfnamefont {J.}~\bibnamefont {Chen}}, \bibinfo {author}
  {\bibfnamefont {Q.}~\bibnamefont {Xu}}, \bibinfo {author} {\bibfnamefont
  {H.}~\bibnamefont {Pang}},\ and\ \bibinfo {author} {\bibfnamefont
  {X.}~\bibnamefont {Hu}},\ }\href@noop {} {\bibfield  {journal} {\bibinfo
  {journal} {ACS Appl. Mater. Interfaces}\ }\textbf {\bibinfo {volume} {10}},\
  \bibinfo {pages} {2360} (\bibinfo {year} {2018})}\BibitemShut {NoStop}%
\bibitem [{\citenamefont {Bykov}\ \emph {et~al.}(2021)\citenamefont {Bykov},
  \citenamefont {Bykova}, \citenamefont {Ponomareva}, \citenamefont
  {Tasn{\'{a}}di}, \citenamefont {Chariton}, \citenamefont {Prakapenka},
  \citenamefont {Glazyrin}, \citenamefont {Smith}, \citenamefont {Mahmood},
  \citenamefont {Abrikosov},\ and\ \citenamefont {Goncharov}}]{Bykov2021}%
  \BibitemOpen
  \bibfield  {author} {\bibinfo {author} {\bibfnamefont {M.}~\bibnamefont
  {Bykov}}, \bibinfo {author} {\bibfnamefont {E.}~\bibnamefont {Bykova}},
  \bibinfo {author} {\bibfnamefont {A.~V.}\ \bibnamefont {Ponomareva}},
  \bibinfo {author} {\bibfnamefont {F.}~\bibnamefont {Tasn{\'{a}}di}}, \bibinfo
  {author} {\bibfnamefont {S.}~\bibnamefont {Chariton}}, \bibinfo {author}
  {\bibfnamefont {V.~B.}\ \bibnamefont {Prakapenka}}, \bibinfo {author}
  {\bibfnamefont {K.}~\bibnamefont {Glazyrin}}, \bibinfo {author}
  {\bibfnamefont {J.~S.}\ \bibnamefont {Smith}}, \bibinfo {author}
  {\bibfnamefont {M.~F.}\ \bibnamefont {Mahmood}}, \bibinfo {author}
  {\bibfnamefont {I.~A.}\ \bibnamefont {Abrikosov}},\ and\ \bibinfo {author}
  {\bibfnamefont {A.~F.}\ \bibnamefont {Goncharov}},\ }\href@noop {} {\bibfield
   {journal} {\bibinfo  {journal} {{ACS} Nano}\ }\textbf {\bibinfo {volume}
  {15}},\ \bibinfo {pages} {13539} (\bibinfo {year} {2021})}\BibitemShut
  {NoStop}%
\bibitem [{\citenamefont {Phan}\ and\ \citenamefont {Chung}(2014)}]{Phan2014}%
  \BibitemOpen
  \bibfield  {author} {\bibinfo {author} {\bibfnamefont {D.-T.}\ \bibnamefont
  {Phan}}\ and\ \bibinfo {author} {\bibfnamefont {G.-S.}\ \bibnamefont
  {Chung}},\ }\href@noop {} {\bibfield  {journal} {\bibinfo  {journal} {Int. J.
  Hydrogen Energy}\ }\textbf {\bibinfo {volume} {39}},\ \bibinfo {pages}
  {20294} (\bibinfo {year} {2014})}\BibitemShut {NoStop}%
\bibitem [{\citenamefont {Ren}\ \emph {et~al.}(2019)\citenamefont {Ren},
  \citenamefont {Mao}, \citenamefont {Luo},\ and\ \citenamefont
  {Liu}}]{Ren2019}%
  \BibitemOpen
  \bibfield  {author} {\bibinfo {author} {\bibfnamefont {Z.}~\bibnamefont
  {Ren}}, \bibinfo {author} {\bibfnamefont {H.}~\bibnamefont {Mao}}, \bibinfo
  {author} {\bibfnamefont {H.}~\bibnamefont {Luo}},\ and\ \bibinfo {author}
  {\bibfnamefont {Y.}~\bibnamefont {Liu}},\ }\href@noop {} {\bibfield
  {journal} {\bibinfo  {journal} {Carbon}\ }\textbf {\bibinfo {volume} {149}},\
  \bibinfo {pages} {609} (\bibinfo {year} {2019})}\BibitemShut {NoStop}%
\bibitem [{\citenamefont {Deokar}\ \emph {et~al.}(2020)\citenamefont {Deokar},
  \citenamefont {Casanova-Ch{\'{a}}fer}, \citenamefont {Rajput}, \citenamefont
  {Aubry}, \citenamefont {Llobet}, \citenamefont {Jouiad},\ and\ \citenamefont
  {Costa}}]{Deokar2020}%
  \BibitemOpen
  \bibfield  {author} {\bibinfo {author} {\bibfnamefont {G.}~\bibnamefont
  {Deokar}}, \bibinfo {author} {\bibfnamefont {J.}~\bibnamefont
  {Casanova-Ch{\'{a}}fer}}, \bibinfo {author} {\bibfnamefont {N.~S.}\
  \bibnamefont {Rajput}}, \bibinfo {author} {\bibfnamefont {C.}~\bibnamefont
  {Aubry}}, \bibinfo {author} {\bibfnamefont {E.}~\bibnamefont {Llobet}},
  \bibinfo {author} {\bibfnamefont {M.}~\bibnamefont {Jouiad}},\ and\ \bibinfo
  {author} {\bibfnamefont {P.~M.}\ \bibnamefont {Costa}},\ }\href@noop {}
  {\bibfield  {journal} {\bibinfo  {journal} {Sens. Actuators B Chem.}\
  }\textbf {\bibinfo {volume} {305}},\ \bibinfo {pages} {127458} (\bibinfo
  {year} {2020})}\BibitemShut {NoStop}%
\bibitem [{\citenamefont {Zhang}\ \emph {et~al.}(2017)\citenamefont {Zhang},
  \citenamefont {Li},\ and\ \citenamefont {Peng}}]{Zhang2017}%
  \BibitemOpen
  \bibfield  {author} {\bibinfo {author} {\bibfnamefont {W.}~\bibnamefont
  {Zhang}}, \bibinfo {author} {\bibfnamefont {Y.}~\bibnamefont {Li}},\ and\
  \bibinfo {author} {\bibfnamefont {S.}~\bibnamefont {Peng}},\ }\href@noop {}
  {\bibfield  {journal} {\bibinfo  {journal} {J. Mater. Chem. A}\ }\textbf
  {\bibinfo {volume} {5}},\ \bibinfo {pages} {13072} (\bibinfo {year}
  {2017})}\BibitemShut {NoStop}%
\bibitem [{\citenamefont {Chen}\ \emph {et~al.}(2019)\citenamefont {Chen},
  \citenamefont {Wang},\ and\ \citenamefont {Liu}}]{Chen2019}%
  \BibitemOpen
  \bibfield  {author} {\bibinfo {author} {\bibfnamefont {D.}~\bibnamefont
  {Chen}}, \bibinfo {author} {\bibfnamefont {W.}~\bibnamefont {Wang}},\ and\
  \bibinfo {author} {\bibfnamefont {C.}~\bibnamefont {Liu}},\ }\href@noop {}
  {\bibfield  {journal} {\bibinfo  {journal} {Int. J. Hydrogen Energy}\
  }\textbf {\bibinfo {volume} {44}},\ \bibinfo {pages} {6560} (\bibinfo {year}
  {2019})}\BibitemShut {NoStop}%
\bibitem [{\citenamefont {Toh}\ \emph {et~al.}(2013)\citenamefont {Toh},
  \citenamefont {Poh}, \citenamefont {Sofer},\ and\ \citenamefont
  {Pumera}}]{Toh2013}%
  \BibitemOpen
  \bibfield  {author} {\bibinfo {author} {\bibfnamefont {R.~J.}\ \bibnamefont
  {Toh}}, \bibinfo {author} {\bibfnamefont {H.~L.}\ \bibnamefont {Poh}},
  \bibinfo {author} {\bibfnamefont {Z.}~\bibnamefont {Sofer}},\ and\ \bibinfo
  {author} {\bibfnamefont {M.}~\bibnamefont {Pumera}},\ }\href@noop {}
  {\bibfield  {journal} {\bibinfo  {journal} {Chem. Asian J.}\ }\textbf
  {\bibinfo {volume} {8}},\ \bibinfo {pages} {1295} (\bibinfo {year}
  {2013})}\BibitemShut {NoStop}%
\bibitem [{\citenamefont {Xu}\ \emph {et~al.}(2016)\citenamefont {Xu},
  \citenamefont {Li}, \citenamefont {Xu}, \citenamefont {Xu},\ and\
  \citenamefont {Zhao}}]{Xu2016}%
  \BibitemOpen
  \bibfield  {author} {\bibinfo {author} {\bibfnamefont {X.-Y.}\ \bibnamefont
  {Xu}}, \bibinfo {author} {\bibfnamefont {J.}~\bibnamefont {Li}}, \bibinfo
  {author} {\bibfnamefont {H.}~\bibnamefont {Xu}}, \bibinfo {author}
  {\bibfnamefont {X.}~\bibnamefont {Xu}},\ and\ \bibinfo {author}
  {\bibfnamefont {C.}~\bibnamefont {Zhao}},\ }\href@noop {} {\bibfield
  {journal} {\bibinfo  {journal} {New J. Chem.}\ }\textbf {\bibinfo {volume}
  {40}},\ \bibinfo {pages} {9361} (\bibinfo {year} {2016})}\BibitemShut
  {NoStop}%
\bibitem [{\citenamefont {Zhou}\ \emph {et~al.}(2016)\citenamefont {Zhou},
  \citenamefont {Szpunar},\ and\ \citenamefont {Cui}}]{Zhou2016}%
  \BibitemOpen
  \bibfield  {author} {\bibinfo {author} {\bibfnamefont {C.}~\bibnamefont
  {Zhou}}, \bibinfo {author} {\bibfnamefont {J.~A.}\ \bibnamefont {Szpunar}},\
  and\ \bibinfo {author} {\bibfnamefont {X.}~\bibnamefont {Cui}},\ }\href@noop
  {} {\bibfield  {journal} {\bibinfo  {journal} {ACS Appl. Mater. Interfaces}\
  }\textbf {\bibinfo {volume} {8}},\ \bibinfo {pages} {15232} (\bibinfo {year}
  {2016})}\BibitemShut {NoStop}%
\bibitem [{\citenamefont {Zhou}\ \emph {et~al.}(2021)\citenamefont {Zhou},
  \citenamefont {Cui}, \citenamefont {Zhou}, \citenamefont {Liu}, \citenamefont
  {Zhao}, \citenamefont {Li},\ and\ \citenamefont {Qu}}]{Zhou2021}%
  \BibitemOpen
  \bibfield  {author} {\bibinfo {author} {\bibfnamefont {D.}~\bibnamefont
  {Zhou}}, \bibinfo {author} {\bibfnamefont {K.}~\bibnamefont {Cui}}, \bibinfo
  {author} {\bibfnamefont {Z.}~\bibnamefont {Zhou}}, \bibinfo {author}
  {\bibfnamefont {C.}~\bibnamefont {Liu}}, \bibinfo {author} {\bibfnamefont
  {W.}~\bibnamefont {Zhao}}, \bibinfo {author} {\bibfnamefont {P.}~\bibnamefont
  {Li}},\ and\ \bibinfo {author} {\bibfnamefont {X.}~\bibnamefont {Qu}},\
  }\href@noop {} {\bibfield  {journal} {\bibinfo  {journal} {Int. J. Hydrogen
  Energy}\ }\textbf {\bibinfo {volume} {46}},\ \bibinfo {pages} {34369}
  (\bibinfo {year} {2021})}\BibitemShut {NoStop}%
\bibitem [{\citenamefont {Safina}\ \emph {et~al.}(2021)\citenamefont {Safina},
  \citenamefont {Krylova}, \citenamefont {Murzaev}, \citenamefont {Baimova},\
  and\ \citenamefont {Mulyukov}}]{Safina2021}%
  \BibitemOpen
  \bibfield  {author} {\bibinfo {author} {\bibfnamefont {L.~R.}\ \bibnamefont
  {Safina}}, \bibinfo {author} {\bibfnamefont {K.~A.}\ \bibnamefont {Krylova}},
  \bibinfo {author} {\bibfnamefont {R.~T.}\ \bibnamefont {Murzaev}}, \bibinfo
  {author} {\bibfnamefont {J.~A.}\ \bibnamefont {Baimova}},\ and\ \bibinfo
  {author} {\bibfnamefont {R.~R.}\ \bibnamefont {Mulyukov}},\ }\href@noop {}
  {\bibfield  {journal} {\bibinfo  {journal} {Materials}\ }\textbf {\bibinfo
  {volume} {14}},\ \bibinfo {pages} {2098} (\bibinfo {year}
  {2021})}\BibitemShut {NoStop}%
\bibitem [{\citenamefont {Manjanath}\ \emph {et~al.}(2014)\citenamefont
  {Manjanath}, \citenamefont {Kumar},\ and\ \citenamefont
  {Singh}}]{Manjanath2014}%
  \BibitemOpen
  \bibfield  {author} {\bibinfo {author} {\bibfnamefont {A.}~\bibnamefont
  {Manjanath}}, \bibinfo {author} {\bibfnamefont {V.}~\bibnamefont {Kumar}},\
  and\ \bibinfo {author} {\bibfnamefont {A.~K.}\ \bibnamefont {Singh}},\
  }\href@noop {} {\bibfield  {journal} {\bibinfo  {journal} {Phys. Chem. Chem.
  Phys.}\ }\textbf {\bibinfo {volume} {16}},\ \bibinfo {pages} {1667} (\bibinfo
  {year} {2014})}\BibitemShut {NoStop}%
\bibitem [{\citenamefont {Li}\ \emph {et~al.}(2018{\natexlab{a}})\citenamefont
  {Li}, \citenamefont {Ren}, \citenamefont {Ao},\ and\ \citenamefont
  {Liu}}]{Li2018}%
  \BibitemOpen
  \bibfield  {author} {\bibinfo {author} {\bibfnamefont {S.}~\bibnamefont
  {Li}}, \bibinfo {author} {\bibfnamefont {J.-C.}\ \bibnamefont {Ren}},
  \bibinfo {author} {\bibfnamefont {Z.}~\bibnamefont {Ao}},\ and\ \bibinfo
  {author} {\bibfnamefont {W.}~\bibnamefont {Liu}},\ }\href@noop {} {\bibfield
  {journal} {\bibinfo  {journal} {Chem. Phys. Lett.}\ }\textbf {\bibinfo
  {volume} {706}},\ \bibinfo {pages} {202} (\bibinfo {year}
  {2018}{\natexlab{a}})}\BibitemShut {NoStop}%
\bibitem [{\citenamefont {Li}\ \emph {et~al.}(2018{\natexlab{b}})\citenamefont
  {Li}, \citenamefont {Zhang}, \citenamefont {Chen}, \citenamefont {Xiao},\
  and\ \citenamefont {Tang}}]{Li2018a}%
  \BibitemOpen
  \bibfield  {author} {\bibinfo {author} {\bibfnamefont {Y.}~\bibnamefont
  {Li}}, \bibinfo {author} {\bibfnamefont {X.}~\bibnamefont {Zhang}}, \bibinfo
  {author} {\bibfnamefont {D.}~\bibnamefont {Chen}}, \bibinfo {author}
  {\bibfnamefont {S.}~\bibnamefont {Xiao}},\ and\ \bibinfo {author}
  {\bibfnamefont {J.}~\bibnamefont {Tang}},\ }\href@noop {} {\bibfield
  {journal} {\bibinfo  {journal} {Appl. Surf. Sci.}\ }\textbf {\bibinfo
  {volume} {443}},\ \bibinfo {pages} {274} (\bibinfo {year}
  {2018}{\natexlab{b}})}\BibitemShut {NoStop}%
\bibitem [{\citenamefont {Mounet}\ \emph {et~al.}(2018)\citenamefont {Mounet},
  \citenamefont {Gibertini}, \citenamefont {Schwaller}, \citenamefont {Campi},
  \citenamefont {Merkys}, \citenamefont {Marrazzo}, \citenamefont {Sohier},
  \citenamefont {Castelli}, \citenamefont {Cepellotti}, \citenamefont {Pizzi},\
  and\ \citenamefont {Marzari}}]{Mounet2018}%
  \BibitemOpen
  \bibfield  {author} {\bibinfo {author} {\bibfnamefont {N.}~\bibnamefont
  {Mounet}}, \bibinfo {author} {\bibfnamefont {M.}~\bibnamefont {Gibertini}},
  \bibinfo {author} {\bibfnamefont {P.}~\bibnamefont {Schwaller}}, \bibinfo
  {author} {\bibfnamefont {D.}~\bibnamefont {Campi}}, \bibinfo {author}
  {\bibfnamefont {A.}~\bibnamefont {Merkys}}, \bibinfo {author} {\bibfnamefont
  {A.}~\bibnamefont {Marrazzo}}, \bibinfo {author} {\bibfnamefont
  {T.}~\bibnamefont {Sohier}}, \bibinfo {author} {\bibfnamefont {I.~E.}\
  \bibnamefont {Castelli}}, \bibinfo {author} {\bibfnamefont {A.}~\bibnamefont
  {Cepellotti}}, \bibinfo {author} {\bibfnamefont {G.}~\bibnamefont {Pizzi}},\
  and\ \bibinfo {author} {\bibfnamefont {N.}~\bibnamefont {Marzari}},\
  }\href@noop {} {\bibfield  {journal} {\bibinfo  {journal} {Nature
  Nanotechnol.}\ }\textbf {\bibinfo {volume} {13}},\ \bibinfo {pages} {246}
  (\bibinfo {year} {2018})}\BibitemShut {NoStop}%
\bibitem [{\citenamefont {McCreary}\ \emph {et~al.}(2014)\citenamefont
  {McCreary}, \citenamefont {Hanbicki}, \citenamefont {Robinson}, \citenamefont
  {Cobas}, \citenamefont {Culbertson}, \citenamefont {Friedman}, \citenamefont
  {Jernigan},\ and\ \citenamefont {Jonker}}]{McCreary2014}%
  \BibitemOpen
  \bibfield  {author} {\bibinfo {author} {\bibfnamefont {K.~M.}\ \bibnamefont
  {McCreary}}, \bibinfo {author} {\bibfnamefont {A.~T.}\ \bibnamefont
  {Hanbicki}}, \bibinfo {author} {\bibfnamefont {J.~T.}\ \bibnamefont
  {Robinson}}, \bibinfo {author} {\bibfnamefont {E.}~\bibnamefont {Cobas}},
  \bibinfo {author} {\bibfnamefont {J.~C.}\ \bibnamefont {Culbertson}},
  \bibinfo {author} {\bibfnamefont {A.~L.}\ \bibnamefont {Friedman}}, \bibinfo
  {author} {\bibfnamefont {G.~G.}\ \bibnamefont {Jernigan}},\ and\ \bibinfo
  {author} {\bibfnamefont {B.~T.}\ \bibnamefont {Jonker}},\ }\href@noop {}
  {\bibfield  {journal} {\bibinfo  {journal} {Adv. Funct. Mater.}\ }\textbf
  {\bibinfo {volume} {24}},\ \bibinfo {pages} {6449} (\bibinfo {year}
  {2014})}\BibitemShut {NoStop}%
\bibitem [{\citenamefont {Wang}\ \emph {et~al.}(2017)\citenamefont {Wang},
  \citenamefont {Huang}, \citenamefont {Lin}, \citenamefont {Cui},
  \citenamefont {Chen}, \citenamefont {Zhu}, \citenamefont {Liu}, \citenamefont
  {Zeng}, \citenamefont {Zhou}, \citenamefont {Yu}, \citenamefont {Wang},
  \citenamefont {He}, \citenamefont {Tsang}, \citenamefont {Gao}, \citenamefont
  {Suenaga}, \citenamefont {Ma}, \citenamefont {Yang}, \citenamefont {Lu},
  \citenamefont {Yu}, \citenamefont {Teo}, \citenamefont {Liu},\ and\
  \citenamefont {Liu}}]{Wang2017}%
  \BibitemOpen
  \bibfield  {author} {\bibinfo {author} {\bibfnamefont {H.}~\bibnamefont
  {Wang}}, \bibinfo {author} {\bibfnamefont {X.}~\bibnamefont {Huang}},
  \bibinfo {author} {\bibfnamefont {J.}~\bibnamefont {Lin}}, \bibinfo {author}
  {\bibfnamefont {J.}~\bibnamefont {Cui}}, \bibinfo {author} {\bibfnamefont
  {Y.}~\bibnamefont {Chen}}, \bibinfo {author} {\bibfnamefont {C.}~\bibnamefont
  {Zhu}}, \bibinfo {author} {\bibfnamefont {F.}~\bibnamefont {Liu}}, \bibinfo
  {author} {\bibfnamefont {Q.}~\bibnamefont {Zeng}}, \bibinfo {author}
  {\bibfnamefont {J.}~\bibnamefont {Zhou}}, \bibinfo {author} {\bibfnamefont
  {P.}~\bibnamefont {Yu}}, \bibinfo {author} {\bibfnamefont {X.}~\bibnamefont
  {Wang}}, \bibinfo {author} {\bibfnamefont {H.}~\bibnamefont {He}}, \bibinfo
  {author} {\bibfnamefont {S.~H.}\ \bibnamefont {Tsang}}, \bibinfo {author}
  {\bibfnamefont {W.}~\bibnamefont {Gao}}, \bibinfo {author} {\bibfnamefont
  {K.}~\bibnamefont {Suenaga}}, \bibinfo {author} {\bibfnamefont
  {F.}~\bibnamefont {Ma}}, \bibinfo {author} {\bibfnamefont {C.}~\bibnamefont
  {Yang}}, \bibinfo {author} {\bibfnamefont {L.}~\bibnamefont {Lu}}, \bibinfo
  {author} {\bibfnamefont {T.}~\bibnamefont {Yu}}, \bibinfo {author}
  {\bibfnamefont {E.~H.~T.}\ \bibnamefont {Teo}}, \bibinfo {author}
  {\bibfnamefont {G.}~\bibnamefont {Liu}},\ and\ \bibinfo {author}
  {\bibfnamefont {Z.}~\bibnamefont {Liu}},\ }\href@noop {} {\bibfield
  {journal} {\bibinfo  {journal} {Nature Commun.}\ }\textbf {\bibinfo {volume}
  {8}},\ \bibinfo {pages} {394} (\bibinfo {year} {2017})}\BibitemShut {NoStop}%
\bibitem [{\citenamefont {Sherrell}\ \emph {et~al.}(2018)\citenamefont
  {Sherrell}, \citenamefont {Sharda}, \citenamefont {Grotta}, \citenamefont
  {Ranalli}, \citenamefont {Sokolikova}, \citenamefont {Pesci}, \citenamefont
  {Palczynski}, \citenamefont {Bemmer},\ and\ \citenamefont
  {Mattevi}}]{Sherrell2018}%
  \BibitemOpen
  \bibfield  {author} {\bibinfo {author} {\bibfnamefont {P.~C.}\ \bibnamefont
  {Sherrell}}, \bibinfo {author} {\bibfnamefont {K.}~\bibnamefont {Sharda}},
  \bibinfo {author} {\bibfnamefont {C.}~\bibnamefont {Grotta}}, \bibinfo
  {author} {\bibfnamefont {J.}~\bibnamefont {Ranalli}}, \bibinfo {author}
  {\bibfnamefont {M.~S.}\ \bibnamefont {Sokolikova}}, \bibinfo {author}
  {\bibfnamefont {F.~M.}\ \bibnamefont {Pesci}}, \bibinfo {author}
  {\bibfnamefont {P.}~\bibnamefont {Palczynski}}, \bibinfo {author}
  {\bibfnamefont {V.~L.}\ \bibnamefont {Bemmer}},\ and\ \bibinfo {author}
  {\bibfnamefont {C.}~\bibnamefont {Mattevi}},\ }\href@noop {} {\bibfield
  {journal} {\bibinfo  {journal} {{ACS} Omega}\ }\textbf {\bibinfo {volume}
  {3}},\ \bibinfo {pages} {8655} (\bibinfo {year} {2018})}\BibitemShut
  {NoStop}%
\bibitem [{\citenamefont {Cong}\ \emph {et~al.}(2013)\citenamefont {Cong},
  \citenamefont {Shang}, \citenamefont {Wu}, \citenamefont {Cao}, \citenamefont
  {Peimyoo}, \citenamefont {Qiu}, \citenamefont {Sun},\ and\ \citenamefont
  {Yu}}]{Cong2013}%
  \BibitemOpen
  \bibfield  {author} {\bibinfo {author} {\bibfnamefont {C.}~\bibnamefont
  {Cong}}, \bibinfo {author} {\bibfnamefont {J.}~\bibnamefont {Shang}},
  \bibinfo {author} {\bibfnamefont {X.}~\bibnamefont {Wu}}, \bibinfo {author}
  {\bibfnamefont {B.}~\bibnamefont {Cao}}, \bibinfo {author} {\bibfnamefont
  {N.}~\bibnamefont {Peimyoo}}, \bibinfo {author} {\bibfnamefont
  {C.}~\bibnamefont {Qiu}}, \bibinfo {author} {\bibfnamefont {L.}~\bibnamefont
  {Sun}},\ and\ \bibinfo {author} {\bibfnamefont {T.}~\bibnamefont {Yu}},\
  }\href@noop {} {\bibfield  {journal} {\bibinfo  {journal} {Adv. Opt. Mater.}\
  }\textbf {\bibinfo {volume} {2}},\ \bibinfo {pages} {131} (\bibinfo {year}
  {2013})}\BibitemShut {NoStop}%
\bibitem [{\citenamefont {Ikeda}\ \emph {et~al.}(2016)\citenamefont {Ikeda},
  \citenamefont {Krockenberger}, \citenamefont {Irie}, \citenamefont {Naito},\
  and\ \citenamefont {Yamamoto}}]{Ikeda2016}%
  \BibitemOpen
  \bibfield  {author} {\bibinfo {author} {\bibfnamefont {A.}~\bibnamefont
  {Ikeda}}, \bibinfo {author} {\bibfnamefont {Y.}~\bibnamefont
  {Krockenberger}}, \bibinfo {author} {\bibfnamefont {H.}~\bibnamefont {Irie}},
  \bibinfo {author} {\bibfnamefont {M.}~\bibnamefont {Naito}},\ and\ \bibinfo
  {author} {\bibfnamefont {H.}~\bibnamefont {Yamamoto}},\ }\href@noop {}
  {\bibfield  {journal} {\bibinfo  {journal} {Appl. Phys. Express}\ }\textbf
  {\bibinfo {volume} {9}},\ \bibinfo {pages} {061101} (\bibinfo {year}
  {2016})}\BibitemShut {NoStop}%
\bibitem [{\citenamefont {Luo}\ \emph {et~al.}(2020)\citenamefont {Luo},
  \citenamefont {Lin}, \citenamefont {Tang}, \citenamefont {Feng},
  \citenamefont {Gui}, \citenamefont {Zhu}, \citenamefont {Yang}, \citenamefont
  {Li}, \citenamefont {Zhou},\ and\ \citenamefont {Fu}}]{Luo2020}%
  \BibitemOpen
  \bibfield  {author} {\bibinfo {author} {\bibfnamefont {Z.}~\bibnamefont
  {Luo}}, \bibinfo {author} {\bibfnamefont {X.}~\bibnamefont {Lin}}, \bibinfo
  {author} {\bibfnamefont {L.}~\bibnamefont {Tang}}, \bibinfo {author}
  {\bibfnamefont {Y.}~\bibnamefont {Feng}}, \bibinfo {author} {\bibfnamefont
  {Y.}~\bibnamefont {Gui}}, \bibinfo {author} {\bibfnamefont {J.}~\bibnamefont
  {Zhu}}, \bibinfo {author} {\bibfnamefont {W.}~\bibnamefont {Yang}}, \bibinfo
  {author} {\bibfnamefont {D.}~\bibnamefont {Li}}, \bibinfo {author}
  {\bibfnamefont {L.}~\bibnamefont {Zhou}},\ and\ \bibinfo {author}
  {\bibfnamefont {L.}~\bibnamefont {Fu}},\ }\href@noop {} {\bibfield  {journal}
  {\bibinfo  {journal} {ACS Appl. Mater. Interfaces}\ }\textbf {\bibinfo
  {volume} {12}},\ \bibinfo {pages} {34755} (\bibinfo {year}
  {2020})}\BibitemShut {NoStop}%
\bibitem [{\citenamefont {Dai}\ \emph {et~al.}(2020)\citenamefont {Dai},
  \citenamefont {Li}, \citenamefont {Li}, \citenamefont {Zhao}, \citenamefont
  {Wu}, \citenamefont {Ma},\ and\ \citenamefont {Duan}}]{Dai2020}%
  \BibitemOpen
  \bibfield  {author} {\bibinfo {author} {\bibfnamefont {C.}~\bibnamefont
  {Dai}}, \bibinfo {author} {\bibfnamefont {B.}~\bibnamefont {Li}}, \bibinfo
  {author} {\bibfnamefont {J.}~\bibnamefont {Li}}, \bibinfo {author}
  {\bibfnamefont {B.}~\bibnamefont {Zhao}}, \bibinfo {author} {\bibfnamefont
  {R.}~\bibnamefont {Wu}}, \bibinfo {author} {\bibfnamefont {H.}~\bibnamefont
  {Ma}},\ and\ \bibinfo {author} {\bibfnamefont {X.}~\bibnamefont {Duan}},\
  }\href@noop {} {\bibfield  {journal} {\bibinfo  {journal} {Nano Res.}\
  }\textbf {\bibinfo {volume} {13}},\ \bibinfo {pages} {2506} (\bibinfo {year}
  {2020})}\BibitemShut {NoStop}%
\bibitem [{\citenamefont {Liu}\ \emph {et~al.}(2018)\citenamefont {Liu},
  \citenamefont {Li}, \citenamefont {Zhang}, \citenamefont {Sun}, \citenamefont
  {Zhou},\ and\ \citenamefont {Song}}]{Liu2018}%
  \BibitemOpen
  \bibfield  {author} {\bibinfo {author} {\bibfnamefont {S.}~\bibnamefont
  {Liu}}, \bibinfo {author} {\bibfnamefont {D.}~\bibnamefont {Li}}, \bibinfo
  {author} {\bibfnamefont {G.}~\bibnamefont {Zhang}}, \bibinfo {author}
  {\bibfnamefont {D.}~\bibnamefont {Sun}}, \bibinfo {author} {\bibfnamefont
  {J.}~\bibnamefont {Zhou}},\ and\ \bibinfo {author} {\bibfnamefont
  {H.}~\bibnamefont {Song}},\ }\href@noop {} {\bibfield  {journal} {\bibinfo
  {journal} {ACS Appl. Mater. Interfaces}\ }\textbf {\bibinfo {volume} {10}},\
  \bibinfo {pages} {34193} (\bibinfo {year} {2018})}\BibitemShut {NoStop}%
\bibitem [{\citenamefont {Solov'yov}\ \emph {et~al.}(2012)\citenamefont
  {Solov'yov}, \citenamefont {Yakubovich}, \citenamefont {Nikolaev},
  \citenamefont {Volkovets},\ and\ \citenamefont {Solov'yov}}]{Solovyov2012}%
  \BibitemOpen
  \bibfield  {author} {\bibinfo {author} {\bibfnamefont {I.~A.}\ \bibnamefont
  {Solov'yov}}, \bibinfo {author} {\bibfnamefont {A.~V.}\ \bibnamefont
  {Yakubovich}}, \bibinfo {author} {\bibfnamefont {P.~V.}\ \bibnamefont
  {Nikolaev}}, \bibinfo {author} {\bibfnamefont {I.}~\bibnamefont
  {Volkovets}},\ and\ \bibinfo {author} {\bibfnamefont {A.~V.}\ \bibnamefont
  {Solov'yov}},\ }\href@noop {} {\bibfield  {journal} {\bibinfo  {journal} {J.
  Comput. Chem.}\ }\textbf {\bibinfo {volume} {33}},\ \bibinfo {pages} {2412}
  (\bibinfo {year} {2012})}\BibitemShut {NoStop}%
\bibitem [{\citenamefont {Sushko}\ \emph {et~al.}(2019)\citenamefont {Sushko},
  \citenamefont {Solov'yov},\ and\ \citenamefont {Solov'yov}}]{Sushko2019}%
  \BibitemOpen
  \bibfield  {author} {\bibinfo {author} {\bibfnamefont {G.~B.}\ \bibnamefont
  {Sushko}}, \bibinfo {author} {\bibfnamefont {I.~A.}\ \bibnamefont
  {Solov'yov}},\ and\ \bibinfo {author} {\bibfnamefont {A.~V.}\ \bibnamefont
  {Solov'yov}},\ }\href@noop {} {\bibfield  {journal} {\bibinfo  {journal} {J.
  Mol. Graph. Model.}\ }\textbf {\bibinfo {volume} {88}},\ \bibinfo {pages}
  {247} (\bibinfo {year} {2019})}\BibitemShut {NoStop}%
\bibitem [{\citenamefont {Lu}\ \emph {et~al.}(2019)\citenamefont {Lu},
  \citenamefont {Yao}, \citenamefont {Xiao}, \citenamefont {Huang},\ and\
  \citenamefont {Kan}}]{Lu_2019_ACSOmega.4.5714}%
  \BibitemOpen
  \bibfield  {author} {\bibinfo {author} {\bibfnamefont {M.}~\bibnamefont
  {Lu}}, \bibinfo {author} {\bibfnamefont {Q.}~\bibnamefont {Yao}}, \bibinfo
  {author} {\bibfnamefont {C.}~\bibnamefont {Xiao}}, \bibinfo {author}
  {\bibfnamefont {C.}~\bibnamefont {Huang}},\ and\ \bibinfo {author}
  {\bibfnamefont {E.}~\bibnamefont {Kan}},\ }\href@noop {} {\bibfield
  {journal} {\bibinfo  {journal} {ACS Omega}\ }\textbf {\bibinfo {volume}
  {4}},\ \bibinfo {pages} {5714} (\bibinfo {year} {2019})}\BibitemShut
  {NoStop}%
\bibitem [{\citenamefont {Kistanov}\ \emph {et~al.}(2022)\citenamefont
  {Kistanov}, \citenamefont {Shcherbinin}, \citenamefont {Botella},
  \citenamefont {Davletshin},\ and\ \citenamefont {Cao}}]{Kistanov_2022_JPCL}%
  \BibitemOpen
  \bibfield  {author} {\bibinfo {author} {\bibfnamefont {A.~A.}\ \bibnamefont
  {Kistanov}}, \bibinfo {author} {\bibfnamefont {S.~A.}\ \bibnamefont
  {Shcherbinin}}, \bibinfo {author} {\bibfnamefont {R.}~\bibnamefont
  {Botella}}, \bibinfo {author} {\bibfnamefont {A.}~\bibnamefont
  {Davletshin}},\ and\ \bibinfo {author} {\bibfnamefont {W.}~\bibnamefont
  {Cao}},\ }\href@noop {} {\bibfield  {journal} {\bibinfo  {journal} {J. Phys.
  Chem. Lett.}\ }\textbf {\bibinfo {volume} {13}},\ \bibinfo {pages} {2165}
  (\bibinfo {year} {2022})}\BibitemShut {NoStop}%
\bibitem [{\citenamefont {Giannozzi}\ \emph {et~al.}(2009)\citenamefont
  {Giannozzi}, \citenamefont {Baroni}, \citenamefont {Bonini}, \citenamefont
  {Calandra}, \citenamefont {Car}, \citenamefont {Cavazzoni}, \citenamefont
  {Ceresoli}, \citenamefont {Chiarotti}, \citenamefont {Cococcioni},
  \citenamefont {Dabo}, \citenamefont {Corso}, \citenamefont {de~Gironcoli},
  \citenamefont {Fabris}, \citenamefont {Fratesi}, \citenamefont {Gebauer},
  \citenamefont {Gerstmann}, \citenamefont {Gougoussis}, \citenamefont
  {Kokalj}, \citenamefont {Lazzeri}, \citenamefont {Martin-Samos},
  \citenamefont {Marzari}, \citenamefont {Mauri}, \citenamefont {Mazzarello},
  \citenamefont {Paolini}, \citenamefont {Pasquarello}, \citenamefont
  {Paulatto}, \citenamefont {Sbraccia}, \citenamefont {Scandolo}, \citenamefont
  {Sclauzero}, \citenamefont {Seitsonen}, \citenamefont {Smogunov},
  \citenamefont {Umari},\ and\ \citenamefont {Wentzcovitch}}]{Giannozzi2009}%
  \BibitemOpen
  \bibfield  {author} {\bibinfo {author} {\bibfnamefont {P.}~\bibnamefont
  {Giannozzi}}, \bibinfo {author} {\bibfnamefont {S.}~\bibnamefont {Baroni}},
  \bibinfo {author} {\bibfnamefont {N.}~\bibnamefont {Bonini}}, \bibinfo
  {author} {\bibfnamefont {M.}~\bibnamefont {Calandra}}, \bibinfo {author}
  {\bibfnamefont {R.}~\bibnamefont {Car}}, \bibinfo {author} {\bibfnamefont
  {C.}~\bibnamefont {Cavazzoni}}, \bibinfo {author} {\bibfnamefont
  {D.}~\bibnamefont {Ceresoli}}, \bibinfo {author} {\bibfnamefont {G.~L.}\
  \bibnamefont {Chiarotti}}, \bibinfo {author} {\bibfnamefont {M.}~\bibnamefont
  {Cococcioni}}, \bibinfo {author} {\bibfnamefont {I.}~\bibnamefont {Dabo}},
  \bibinfo {author} {\bibfnamefont {A.~D.}\ \bibnamefont {Corso}}, \bibinfo
  {author} {\bibfnamefont {S.}~\bibnamefont {de~Gironcoli}}, \bibinfo {author}
  {\bibfnamefont {S.}~\bibnamefont {Fabris}}, \bibinfo {author} {\bibfnamefont
  {G.}~\bibnamefont {Fratesi}}, \bibinfo {author} {\bibfnamefont
  {R.}~\bibnamefont {Gebauer}}, \bibinfo {author} {\bibfnamefont
  {U.}~\bibnamefont {Gerstmann}}, \bibinfo {author} {\bibfnamefont
  {C.}~\bibnamefont {Gougoussis}}, \bibinfo {author} {\bibfnamefont
  {A.}~\bibnamefont {Kokalj}}, \bibinfo {author} {\bibfnamefont
  {M.}~\bibnamefont {Lazzeri}}, \bibinfo {author} {\bibfnamefont
  {L.}~\bibnamefont {Martin-Samos}}, \bibinfo {author} {\bibfnamefont
  {N.}~\bibnamefont {Marzari}}, \bibinfo {author} {\bibfnamefont
  {F.}~\bibnamefont {Mauri}}, \bibinfo {author} {\bibfnamefont
  {R.}~\bibnamefont {Mazzarello}}, \bibinfo {author} {\bibfnamefont
  {S.}~\bibnamefont {Paolini}}, \bibinfo {author} {\bibfnamefont
  {A.}~\bibnamefont {Pasquarello}}, \bibinfo {author} {\bibfnamefont
  {L.}~\bibnamefont {Paulatto}}, \bibinfo {author} {\bibfnamefont
  {C.}~\bibnamefont {Sbraccia}}, \bibinfo {author} {\bibfnamefont
  {S.}~\bibnamefont {Scandolo}}, \bibinfo {author} {\bibfnamefont
  {G.}~\bibnamefont {Sclauzero}}, \bibinfo {author} {\bibfnamefont {A.~P.}\
  \bibnamefont {Seitsonen}}, \bibinfo {author} {\bibfnamefont {A.}~\bibnamefont
  {Smogunov}}, \bibinfo {author} {\bibfnamefont {P.}~\bibnamefont {Umari}},\
  and\ \bibinfo {author} {\bibfnamefont {R.~M.}\ \bibnamefont {Wentzcovitch}},\
  }\href@noop {} {\bibfield  {journal} {\bibinfo  {journal} {J. Phys.: Condens.
  Matter}\ }\textbf {\bibinfo {volume} {21}},\ \bibinfo {pages} {395502}
  (\bibinfo {year} {2009})}\BibitemShut {NoStop}%
\bibitem [{\citenamefont {Giannozzi}\ \emph {et~al.}(2017)\citenamefont
  {Giannozzi}, \citenamefont {Andreussi}, \citenamefont {Brumme}, \citenamefont
  {Bunau}, \citenamefont {Nardelli}, \citenamefont {Calandra}, \citenamefont
  {Car}, \citenamefont {Cavazzoni}, \citenamefont {Ceresoli}, \citenamefont
  {Cococcioni}, \citenamefont {Colonna}, \citenamefont {Carnimeo},
  \citenamefont {Corso}, \citenamefont {de~Gironcoli}, \citenamefont {Delugas},
  \citenamefont {DiStasio}, \citenamefont {Ferretti}, \citenamefont {Floris},
  \citenamefont {Fratesi}, \citenamefont {Fugallo}, \citenamefont {Gebauer},
  \citenamefont {Gerstmann}, \citenamefont {Giustino}, \citenamefont {Gorni},
  \citenamefont {Jia}, \citenamefont {Kawamura}, \citenamefont {Ko},
  \citenamefont {Kokalj}, \citenamefont {Kü{\c{c}}ükbenli}, \citenamefont
  {Lazzeri}, \citenamefont {Marsili}, \citenamefont {Marzari}, \citenamefont
  {Mauri}, \citenamefont {Nguyen}, \citenamefont {Nguyen}, \citenamefont {de-la
  Roza}, \citenamefont {Paulatto}, \citenamefont {Ponc{\'{e}}}, \citenamefont
  {Rocca}, \citenamefont {Sabatini}, \citenamefont {Santra}, \citenamefont
  {Schlipf}, \citenamefont {Seitsonen}, \citenamefont {Smogunov}, \citenamefont
  {Timrov}, \citenamefont {Thonhauser}, \citenamefont {Umari}, \citenamefont
  {Vast}, \citenamefont {Wu},\ and\ \citenamefont {Baroni}}]{Giannozzi2017}%
  \BibitemOpen
  \bibfield  {author} {\bibinfo {author} {\bibfnamefont {P.}~\bibnamefont
  {Giannozzi}}, \bibinfo {author} {\bibfnamefont {O.}~\bibnamefont
  {Andreussi}}, \bibinfo {author} {\bibfnamefont {T.}~\bibnamefont {Brumme}},
  \bibinfo {author} {\bibfnamefont {O.}~\bibnamefont {Bunau}}, \bibinfo
  {author} {\bibfnamefont {M.~B.}\ \bibnamefont {Nardelli}}, \bibinfo {author}
  {\bibfnamefont {M.}~\bibnamefont {Calandra}}, \bibinfo {author}
  {\bibfnamefont {R.}~\bibnamefont {Car}}, \bibinfo {author} {\bibfnamefont
  {C.}~\bibnamefont {Cavazzoni}}, \bibinfo {author} {\bibfnamefont
  {D.}~\bibnamefont {Ceresoli}}, \bibinfo {author} {\bibfnamefont
  {M.}~\bibnamefont {Cococcioni}}, \bibinfo {author} {\bibfnamefont
  {N.}~\bibnamefont {Colonna}}, \bibinfo {author} {\bibfnamefont
  {I.}~\bibnamefont {Carnimeo}}, \bibinfo {author} {\bibfnamefont {A.~D.}\
  \bibnamefont {Corso}}, \bibinfo {author} {\bibfnamefont {S.}~\bibnamefont
  {de~Gironcoli}}, \bibinfo {author} {\bibfnamefont {P.}~\bibnamefont
  {Delugas}}, \bibinfo {author} {\bibfnamefont {R.~A.}\ \bibnamefont
  {DiStasio}}, \bibinfo {author} {\bibfnamefont {A.}~\bibnamefont {Ferretti}},
  \bibinfo {author} {\bibfnamefont {A.}~\bibnamefont {Floris}}, \bibinfo
  {author} {\bibfnamefont {G.}~\bibnamefont {Fratesi}}, \bibinfo {author}
  {\bibfnamefont {G.}~\bibnamefont {Fugallo}}, \bibinfo {author} {\bibfnamefont
  {R.}~\bibnamefont {Gebauer}}, \bibinfo {author} {\bibfnamefont
  {U.}~\bibnamefont {Gerstmann}}, \bibinfo {author} {\bibfnamefont
  {F.}~\bibnamefont {Giustino}}, \bibinfo {author} {\bibfnamefont
  {T.}~\bibnamefont {Gorni}}, \bibinfo {author} {\bibfnamefont
  {J.}~\bibnamefont {Jia}}, \bibinfo {author} {\bibfnamefont {M.}~\bibnamefont
  {Kawamura}}, \bibinfo {author} {\bibfnamefont {H.-Y.}\ \bibnamefont {Ko}},
  \bibinfo {author} {\bibfnamefont {A.}~\bibnamefont {Kokalj}}, \bibinfo
  {author} {\bibfnamefont {E.}~\bibnamefont {Kü{\c{c}}ükbenli}}, \bibinfo
  {author} {\bibfnamefont {M.}~\bibnamefont {Lazzeri}}, \bibinfo {author}
  {\bibfnamefont {M.}~\bibnamefont {Marsili}}, \bibinfo {author} {\bibfnamefont
  {N.}~\bibnamefont {Marzari}}, \bibinfo {author} {\bibfnamefont
  {F.}~\bibnamefont {Mauri}}, \bibinfo {author} {\bibfnamefont {N.~L.}\
  \bibnamefont {Nguyen}}, \bibinfo {author} {\bibfnamefont {H.-V.}\
  \bibnamefont {Nguyen}}, \bibinfo {author} {\bibfnamefont {A.~O.}\
  \bibnamefont {de-la Roza}}, \bibinfo {author} {\bibfnamefont
  {L.}~\bibnamefont {Paulatto}}, \bibinfo {author} {\bibfnamefont
  {S.}~\bibnamefont {Ponc{\'{e}}}}, \bibinfo {author} {\bibfnamefont
  {D.}~\bibnamefont {Rocca}}, \bibinfo {author} {\bibfnamefont
  {R.}~\bibnamefont {Sabatini}}, \bibinfo {author} {\bibfnamefont
  {B.}~\bibnamefont {Santra}}, \bibinfo {author} {\bibfnamefont
  {M.}~\bibnamefont {Schlipf}}, \bibinfo {author} {\bibfnamefont {A.~P.}\
  \bibnamefont {Seitsonen}}, \bibinfo {author} {\bibfnamefont {A.}~\bibnamefont
  {Smogunov}}, \bibinfo {author} {\bibfnamefont {I.}~\bibnamefont {Timrov}},
  \bibinfo {author} {\bibfnamefont {T.}~\bibnamefont {Thonhauser}}, \bibinfo
  {author} {\bibfnamefont {P.}~\bibnamefont {Umari}}, \bibinfo {author}
  {\bibfnamefont {N.}~\bibnamefont {Vast}}, \bibinfo {author} {\bibfnamefont
  {X.}~\bibnamefont {Wu}},\ and\ \bibinfo {author} {\bibfnamefont
  {S.}~\bibnamefont {Baroni}},\ }\href@noop {} {\bibfield  {journal} {\bibinfo
  {journal} {J. Phys.: Condens. Matter}\ }\textbf {\bibinfo {volume} {29}},\
  \bibinfo {pages} {465901} (\bibinfo {year} {2017})}\BibitemShut {NoStop}%
\bibitem [{\citenamefont {Perdew}\ \emph {et~al.}(1996)\citenamefont {Perdew},
  \citenamefont {Burke},\ and\ \citenamefont {Ernzerhof}}]{Perdew1996}%
  \BibitemOpen
  \bibfield  {author} {\bibinfo {author} {\bibfnamefont {J.~P.}\ \bibnamefont
  {Perdew}}, \bibinfo {author} {\bibfnamefont {K.}~\bibnamefont {Burke}},\ and\
  \bibinfo {author} {\bibfnamefont {M.}~\bibnamefont {Ernzerhof}},\ }\href@noop
  {} {\bibfield  {journal} {\bibinfo  {journal} {Phys. Rev. Lett.}\ }\textbf
  {\bibinfo {volume} {77}},\ \bibinfo {pages} {3865} (\bibinfo {year}
  {1996})}\BibitemShut {NoStop}%
\bibitem [{\citenamefont {Bl\"ochl}(1994)}]{Bloechl1994}%
  \BibitemOpen
  \bibfield  {author} {\bibinfo {author} {\bibfnamefont {P.~E.}\ \bibnamefont
  {Bl\"ochl}},\ }\href@noop {} {\bibfield  {journal} {\bibinfo  {journal}
  {Phys. Rev. B}\ }\textbf {\bibinfo {volume} {50}},\ \bibinfo {pages} {17953}
  (\bibinfo {year} {1994})}\BibitemShut {NoStop}%
\bibitem [{\citenamefont {Kresse}\ and\ \citenamefont
  {Joubert}(1999)}]{Kresse1999}%
  \BibitemOpen
  \bibfield  {author} {\bibinfo {author} {\bibfnamefont {G.}~\bibnamefont
  {Kresse}}\ and\ \bibinfo {author} {\bibfnamefont {D.}~\bibnamefont
  {Joubert}},\ }\href@noop {} {\bibfield  {journal} {\bibinfo  {journal} {Phys.
  Rev. B}\ }\textbf {\bibinfo {volume} {59}},\ \bibinfo {pages} {1758}
  (\bibinfo {year} {1999})}\BibitemShut {NoStop}%
\bibitem [{\citenamefont {Grimme}\ \emph {et~al.}(2010)\citenamefont {Grimme},
  \citenamefont {Antony}, \citenamefont {Ehrlich},\ and\ \citenamefont
  {Krieg}}]{Grimme2010}%
  \BibitemOpen
  \bibfield  {author} {\bibinfo {author} {\bibfnamefont {S.}~\bibnamefont
  {Grimme}}, \bibinfo {author} {\bibfnamefont {J.}~\bibnamefont {Antony}},
  \bibinfo {author} {\bibfnamefont {S.}~\bibnamefont {Ehrlich}},\ and\ \bibinfo
  {author} {\bibfnamefont {H.}~\bibnamefont {Krieg}},\ }\href@noop {}
  {\bibfield  {journal} {\bibinfo  {journal} {J. Chem. Phys.}\ }\textbf
  {\bibinfo {volume} {132}},\ \bibinfo {pages} {154104} (\bibinfo {year}
  {2010})}\BibitemShut {NoStop}%
\bibitem [{\citenamefont {Monkhorst}\ and\ \citenamefont
  {Pack}(1976)}]{Monkhorst1976}%
  \BibitemOpen
  \bibfield  {author} {\bibinfo {author} {\bibfnamefont {H.~J.}\ \bibnamefont
  {Monkhorst}}\ and\ \bibinfo {author} {\bibfnamefont {J.~D.}\ \bibnamefont
  {Pack}},\ }\href@noop {} {\bibfield  {journal} {\bibinfo  {journal} {Phys.
  Rev. B}\ }\textbf {\bibinfo {volume} {13}},\ \bibinfo {pages} {5188}
  (\bibinfo {year} {1976})}\BibitemShut {NoStop}%
\bibitem [{\citenamefont {Methfessel}\ and\ \citenamefont
  {Paxton}(1989)}]{Methfessel1989}%
  \BibitemOpen
  \bibfield  {author} {\bibinfo {author} {\bibfnamefont {M.}~\bibnamefont
  {Methfessel}}\ and\ \bibinfo {author} {\bibfnamefont {A.~T.}\ \bibnamefont
  {Paxton}},\ }\href@noop {} {\bibfield  {journal} {\bibinfo  {journal} {Phys.
  Rev. B}\ }\textbf {\bibinfo {volume} {40}},\ \bibinfo {pages} {3616}
  (\bibinfo {year} {1989})}\BibitemShut {NoStop}%
\bibitem [{\citenamefont {Solov'yov}\ \emph
  {et~al.}(2017{\natexlab{a}})\citenamefont {Solov'yov}, \citenamefont
  {Korol},\ and\ \citenamefont {Solov'yov}}]{MBNbook_Springer_2017}%
  \BibitemOpen
  \bibfield  {author} {\bibinfo {author} {\bibfnamefont {I.~A.}\ \bibnamefont
  {Solov'yov}}, \bibinfo {author} {\bibfnamefont {A.~V.}\ \bibnamefont
  {Korol}},\ and\ \bibinfo {author} {\bibfnamefont {A.~V.}\ \bibnamefont
  {Solov'yov}},\ }\href@noop {} {\emph {\bibinfo {title} {{Multiscale Modeling
  of Complex Molecular Structure and Dynamics with MBN Explorer}}}}\ (\bibinfo
  {publisher} {Springer International Publishing},\ \bibinfo {address} {Cham,
  Switzerland},\ \bibinfo {year} {2017})\BibitemShut {NoStop}%
\bibitem [{\citenamefont {Solov'yov}\ \emph {et~al.}(2022)\citenamefont
  {Solov'yov}, \citenamefont {Verkhovtsev}, \citenamefont {Korol},\ and\
  \citenamefont {Solov'yov}}]{DySoN_book_Springer_2022}%
  \BibitemOpen
  \bibinfo {editor} {\bibfnamefont {I.~A.}\ \bibnamefont {Solov'yov}}, \bibinfo
  {editor} {\bibfnamefont {A.~V.}\ \bibnamefont {Verkhovtsev}}, \bibinfo
  {editor} {\bibfnamefont {A.~V.}\ \bibnamefont {Korol}},\ and\ \bibinfo
  {editor} {\bibfnamefont {A.~V.}\ \bibnamefont {Solov'yov}},\ eds.,\
  \href@noop {} {\emph {\bibinfo {title} {{ Dynamics of Systems on the
  Nanoscale}}}}\ (\bibinfo  {publisher} {Springer International Publishing},\
  \bibinfo {address} {Cham, Switzerland},\ \bibinfo {year} {2022})\BibitemShut
  {NoStop}%
\bibitem [{\citenamefont {Verkhovtsev}\ \emph {et~al.}(2014)\citenamefont
  {Verkhovtsev}, \citenamefont {Schramm},\ and\ \citenamefont
  {Solov'yov}}]{Verkhovtsev_2014_CNT}%
  \BibitemOpen
  \bibfield  {author} {\bibinfo {author} {\bibfnamefont {A.~V.}\ \bibnamefont
  {Verkhovtsev}}, \bibinfo {author} {\bibfnamefont {S.}~\bibnamefont
  {Schramm}},\ and\ \bibinfo {author} {\bibfnamefont {A.~V.}\ \bibnamefont
  {Solov'yov}},\ }\href@noop {} {\bibfield  {journal} {\bibinfo  {journal}
  {Eur. Phys. J. D}\ }\textbf {\bibinfo {volume} {68}},\ \bibinfo {pages} {246}
  (\bibinfo {year} {2014})}\BibitemShut {NoStop}%
\bibitem [{\citenamefont {Sushko}\ \emph {et~al.}(2014)\citenamefont {Sushko},
  \citenamefont {Verkhovtsev},\ and\ \citenamefont
  {Solov'yov}}]{Sushko_2014_JPCA_FF}%
  \BibitemOpen
  \bibfield  {author} {\bibinfo {author} {\bibfnamefont {G.~B.}\ \bibnamefont
  {Sushko}}, \bibinfo {author} {\bibfnamefont {A.~V.}\ \bibnamefont
  {Verkhovtsev}},\ and\ \bibinfo {author} {\bibfnamefont {A.~V.}\ \bibnamefont
  {Solov'yov}},\ }\href@noop {} {\bibfield  {journal} {\bibinfo  {journal} {J.
  Phys. Chem. A}\ }\textbf {\bibinfo {volume} {118}},\ \bibinfo {pages} {8426}
  (\bibinfo {year} {2014})}\BibitemShut {NoStop}%
\bibitem [{\citenamefont {Sushko}\ \emph {et~al.}(2016)\citenamefont {Sushko},
  \citenamefont {Solov'yov},\ and\ \citenamefont
  {Solov'yov}}]{Sushko_2016_IDMD}%
  \BibitemOpen
  \bibfield  {author} {\bibinfo {author} {\bibfnamefont {G.~B.}\ \bibnamefont
  {Sushko}}, \bibinfo {author} {\bibfnamefont {I.~A.}\ \bibnamefont
  {Solov'yov}},\ and\ \bibinfo {author} {\bibfnamefont {A.~V.}\ \bibnamefont
  {Solov'yov}},\ }\href@noop {} {\bibfield  {journal} {\bibinfo  {journal}
  {Eur. Phys. J. D}\ }\textbf {\bibinfo {volume} {70}},\ \bibinfo {pages} {217}
  (\bibinfo {year} {2016})}\BibitemShut {NoStop}%
\bibitem [{\citenamefont {de~Vera}\ \emph {et~al.}(2020)\citenamefont
  {de~Vera}, \citenamefont {Azzolini}, \citenamefont {Sushko}, \citenamefont
  {Abril}, \citenamefont {Garcia-Molina}, \citenamefont {Dapor}, \citenamefont
  {Solov'yov},\ and\ \citenamefont {Solov'yov}}]{deVera_2020_FEBID}%
  \BibitemOpen
  \bibfield  {author} {\bibinfo {author} {\bibfnamefont {P.}~\bibnamefont
  {de~Vera}}, \bibinfo {author} {\bibfnamefont {M.}~\bibnamefont {Azzolini}},
  \bibinfo {author} {\bibfnamefont {G.}~\bibnamefont {Sushko}}, \bibinfo
  {author} {\bibfnamefont {I.}~\bibnamefont {Abril}}, \bibinfo {author}
  {\bibfnamefont {R.}~\bibnamefont {Garcia-Molina}}, \bibinfo {author}
  {\bibfnamefont {M.}~\bibnamefont {Dapor}}, \bibinfo {author} {\bibfnamefont
  {I.~A.}\ \bibnamefont {Solov'yov}},\ and\ \bibinfo {author} {\bibfnamefont
  {A.~V.}\ \bibnamefont {Solov'yov}},\ }\href@noop {} {\bibfield  {journal}
  {\bibinfo  {journal} {Sci. Rep.}\ }\textbf {\bibinfo {volume} {10}},\
  \bibinfo {pages} {20827} (\bibinfo {year} {2020})}\BibitemShut {NoStop}%
\bibitem [{\citenamefont {Geng}\ \emph {et~al.}(2009)\citenamefont {Geng},
  \citenamefont {Solov'yov}, \citenamefont {Zhou}, \citenamefont {Solov'yov},\
  and\ \citenamefont {Johnson}}]{Geng_2009_C60-TMB_JPCC}%
  \BibitemOpen
  \bibfield  {author} {\bibinfo {author} {\bibfnamefont {J.}~\bibnamefont
  {Geng}}, \bibinfo {author} {\bibfnamefont {I.~A.}\ \bibnamefont {Solov'yov}},
  \bibinfo {author} {\bibfnamefont {W.}~\bibnamefont {Zhou}}, \bibinfo {author}
  {\bibfnamefont {A.~V.}\ \bibnamefont {Solov'yov}},\ and\ \bibinfo {author}
  {\bibfnamefont {B.~F.~G.}\ \bibnamefont {Johnson}},\ }\href@noop {}
  {\bibfield  {journal} {\bibinfo  {journal} {J. Phys. Chem. C}\ }\textbf
  {\bibinfo {volume} {113}},\ \bibinfo {pages} {6390} (\bibinfo {year}
  {2009})}\BibitemShut {NoStop}%
\bibitem [{\citenamefont {Moskovkin}\ \emph {et~al.}(2014)\citenamefont
  {Moskovkin}, \citenamefont {Panshenskov}, \citenamefont {Lucas},\ and\
  \citenamefont {Solov'yov}}]{Moskovkin_2014}%
  \BibitemOpen
  \bibfield  {author} {\bibinfo {author} {\bibfnamefont {P.}~\bibnamefont
  {Moskovkin}}, \bibinfo {author} {\bibfnamefont {M.}~\bibnamefont
  {Panshenskov}}, \bibinfo {author} {\bibfnamefont {S.}~\bibnamefont {Lucas}},\
  and\ \bibinfo {author} {\bibfnamefont {A.~V.}\ \bibnamefont {Solov'yov}},\
  }\href@noop {} {\bibfield  {journal} {\bibinfo  {journal} {Phys. Stat. Sol.
  B}\ }\textbf {\bibinfo {volume} {251}},\ \bibinfo {pages} {1456} (\bibinfo
  {year} {2014})}\BibitemShut {NoStop}%
\bibitem [{\citenamefont {Atkinson}\ \emph {et~al.}(2002)\citenamefont
  {Atkinson}, \citenamefont {Grimes},\ and\ \citenamefont {Owens}}]{nif2}%
  \BibitemOpen
  \bibfield  {author} {\bibinfo {author} {\bibfnamefont {K.}~\bibnamefont
  {Atkinson}}, \bibinfo {author} {\bibfnamefont {R.}~\bibnamefont {Grimes}},\
  and\ \bibinfo {author} {\bibfnamefont {S.}~\bibnamefont {Owens}},\
  }\href@noop {} {\bibfield  {journal} {\bibinfo  {journal} {Solid State
  Ionics}\ }\textbf {\bibinfo {volume} {150}},\ \bibinfo {pages} {443}
  (\bibinfo {year} {2002})}\BibitemShut {NoStop}%
\bibitem [{\citenamefont {Wilson}\ \emph {et~al.}(2002)\citenamefont {Wilson},
  \citenamefont {Hutchinson},\ and\ \citenamefont {Madden}}]{zncl2}%
  \BibitemOpen
  \bibfield  {author} {\bibinfo {author} {\bibfnamefont {M.}~\bibnamefont
  {Wilson}}, \bibinfo {author} {\bibfnamefont {F.}~\bibnamefont {Hutchinson}},\
  and\ \bibinfo {author} {\bibfnamefont {P.~A.}\ \bibnamefont {Madden}},\
  }\href@noop {} {\bibfield  {journal} {\bibinfo  {journal} {Phys. Rev. B}\
  }\textbf {\bibinfo {volume} {65}},\ \bibinfo {pages} {094109} (\bibinfo
  {year} {2002})}\BibitemShut {NoStop}%
\bibitem [{\citenamefont {Hutchinson}\ \emph {et~al.}(1999)\citenamefont
  {Hutchinson}, \citenamefont {Walters}, \citenamefont {Rowley},\ and\
  \citenamefont {Madden}}]{alcl3}%
  \BibitemOpen
  \bibfield  {author} {\bibinfo {author} {\bibfnamefont {F.}~\bibnamefont
  {Hutchinson}}, \bibinfo {author} {\bibfnamefont {M.~K.}\ \bibnamefont
  {Walters}}, \bibinfo {author} {\bibfnamefont {A.~J.}\ \bibnamefont
  {Rowley}},\ and\ \bibinfo {author} {\bibfnamefont {P.~A.}\ \bibnamefont
  {Madden}},\ }\href@noop {} {\bibfield  {journal} {\bibinfo  {journal} {J.
  Chem. Phys.}\ }\textbf {\bibinfo {volume} {110}},\ \bibinfo {pages} {5821}
  (\bibinfo {year} {1999})}\BibitemShut {NoStop}%
\bibitem [{\citenamefont {Matsui}\ and\ \citenamefont
  {Akaogi}(1991)}]{Matsui_1991}%
  \BibitemOpen
  \bibfield  {author} {\bibinfo {author} {\bibfnamefont {M.}~\bibnamefont
  {Matsui}}\ and\ \bibinfo {author} {\bibfnamefont {M.}~\bibnamefont
  {Akaogi}},\ }\href@noop {} {\bibfield  {journal} {\bibinfo  {journal} {Molec.
  Simul.}\ }\textbf {\bibinfo {volume} {6}},\ \bibinfo {pages} {239} (\bibinfo
  {year} {1991})}\BibitemShut {NoStop}%
\bibitem [{\citenamefont {Mattoni}\ \emph {et~al.}(2015)\citenamefont
  {Mattoni}, \citenamefont {Filippetti}, \citenamefont {Saba},\ and\
  \citenamefont {Delugas}}]{Mattoni_2015_JPCC.119.17421}%
  \BibitemOpen
  \bibfield  {author} {\bibinfo {author} {\bibfnamefont {A.}~\bibnamefont
  {Mattoni}}, \bibinfo {author} {\bibfnamefont {A.}~\bibnamefont {Filippetti}},
  \bibinfo {author} {\bibfnamefont {M.~I.}\ \bibnamefont {Saba}},\ and\
  \bibinfo {author} {\bibfnamefont {P.}~\bibnamefont {Delugas}},\ }\href@noop
  {} {\bibfield  {journal} {\bibinfo  {journal} {J. Phys. Chem. C}\ }\textbf
  {\bibinfo {volume} {119}},\ \bibinfo {pages} {17421} (\bibinfo {year}
  {2015})}\BibitemShut {NoStop}%
\bibitem [{\citenamefont {Morgan}\ \emph {et~al.}(2018)\citenamefont {Morgan},
  \citenamefont {Molinari}, \citenamefont {Corrias},\ and\ \citenamefont
  {Sayle}}]{Morgan_2018}%
  \BibitemOpen
  \bibfield  {author} {\bibinfo {author} {\bibfnamefont {L.~M.}\ \bibnamefont
  {Morgan}}, \bibinfo {author} {\bibfnamefont {M.}~\bibnamefont {Molinari}},
  \bibinfo {author} {\bibfnamefont {A.}~\bibnamefont {Corrias}},\ and\ \bibinfo
  {author} {\bibfnamefont {D.~C.}\ \bibnamefont {Sayle}},\ }\href@noop {}
  {\bibfield  {journal} {\bibinfo  {journal} {ACS Appl. Mater. Interfaces}\
  }\textbf {\bibinfo {volume} {10}},\ \bibinfo {pages} {32510} (\bibinfo {year}
  {2018})}\BibitemShut {NoStop}%
\bibitem [{\citenamefont {Pedone}\ \emph {et~al.}(2022)\citenamefont {Pedone},
  \citenamefont {Bertani}, \citenamefont {Brugnoli},\ and\ \citenamefont
  {Pallini}}]{Pedone_2022}%
  \BibitemOpen
  \bibfield  {author} {\bibinfo {author} {\bibfnamefont {A.}~\bibnamefont
  {Pedone}}, \bibinfo {author} {\bibfnamefont {M.}~\bibnamefont {Bertani}},
  \bibinfo {author} {\bibfnamefont {L.}~\bibnamefont {Brugnoli}},\ and\
  \bibinfo {author} {\bibfnamefont {A.}~\bibnamefont {Pallini}},\ }\href@noop
  {} {\bibfield  {journal} {\bibinfo  {journal} {J. Non-Cryst. Solids: X}\
  }\textbf {\bibinfo {volume} {15}},\ \bibinfo {pages} {100115} (\bibinfo
  {year} {2022})}\BibitemShut {NoStop}%
\bibitem [{\citenamefont {Nelder}\ and\ \citenamefont
  {Mead}(1965)}]{nelder-mead}%
  \BibitemOpen
  \bibfield  {author} {\bibinfo {author} {\bibfnamefont {J.~A.}\ \bibnamefont
  {Nelder}}\ and\ \bibinfo {author} {\bibfnamefont {R.}~\bibnamefont {Mead}},\
  }\href@noop {} {\bibfield  {journal} {\bibinfo  {journal} {Comput. J.}\
  }\textbf {\bibinfo {volume} {7}},\ \bibinfo {pages} {308} (\bibinfo {year}
  {1965})}\BibitemShut {NoStop}%
\bibitem [{\citenamefont {Gao}\ and\ \citenamefont {Han}(2012)}]{python}%
  \BibitemOpen
  \bibfield  {author} {\bibinfo {author} {\bibfnamefont {F.}~\bibnamefont
  {Gao}}\ and\ \bibinfo {author} {\bibfnamefont {L.}~\bibnamefont {Han}},\
  }\href {http://dblp.uni-trier.de/db/journals/coap/coap51.html#GaoH12}
  {\bibfield  {journal} {\bibinfo  {journal} {Comput. Optim. Appl.}\ }\textbf
  {\bibinfo {volume} {51}},\ \bibinfo {pages} {259} (\bibinfo {year}
  {2012})}\BibitemShut {NoStop}%
\bibitem [{\citenamefont {Li}\ \emph {et~al.}(2015)\citenamefont {Li},
  \citenamefont {Song},\ and\ \citenamefont {{Merz,
  Jr.}}}]{Li_2015_JPCB.119.883}%
  \BibitemOpen
  \bibfield  {author} {\bibinfo {author} {\bibfnamefont {P.}~\bibnamefont
  {Li}}, \bibinfo {author} {\bibfnamefont {L.~F.}\ \bibnamefont {Song}},\ and\
  \bibinfo {author} {\bibfnamefont {K.~M.}\ \bibnamefont {{Merz, Jr.}}},\
  }\href@noop {} {\bibfield  {journal} {\bibinfo  {journal} {J. Phys. Chem. B}\
  }\textbf {\bibinfo {volume} {119}},\ \bibinfo {pages} {883} (\bibinfo {year}
  {2015})}\BibitemShut {NoStop}%
\bibitem [{\citenamefont {Turupcu}\ \emph {et~al.}(2020)\citenamefont
  {Turupcu}, \citenamefont {Tirado-Rives},\ and\ \citenamefont
  {Jorgensen}}]{Turupcu_2020_JCTC.16.7184}%
  \BibitemOpen
  \bibfield  {author} {\bibinfo {author} {\bibfnamefont {A.}~\bibnamefont
  {Turupcu}}, \bibinfo {author} {\bibfnamefont {J.}~\bibnamefont
  {Tirado-Rives}},\ and\ \bibinfo {author} {\bibfnamefont {W.~L.}\ \bibnamefont
  {Jorgensen}},\ }\href@noop {} {\bibfield  {journal} {\bibinfo  {journal} {J.
  Chem. Theory Comput.}\ }\textbf {\bibinfo {volume} {16}},\ \bibinfo {pages}
  {7184} (\bibinfo {year} {2020})}\BibitemShut {NoStop}%
\bibitem [{\citenamefont {Li}\ and\ \citenamefont {{Merz
  Jr.}}(2017)}]{Li_2017_ChemRev.117.1564}%
  \BibitemOpen
  \bibfield  {author} {\bibinfo {author} {\bibfnamefont {P.}~\bibnamefont
  {Li}}\ and\ \bibinfo {author} {\bibfnamefont {K.~M.}\ \bibnamefont {{Merz
  Jr.}}},\ }\href@noop {} {\bibfield  {journal} {\bibinfo  {journal} {Chem.
  Rev.}\ }\textbf {\bibinfo {volume} {117}},\ \bibinfo {pages} {1564} (\bibinfo
  {year} {2017})}\BibitemShut {NoStop}%
\bibitem [{\citenamefont {Mason}\ \emph {et~al.}(1972)\citenamefont {Mason},
  \citenamefont {{O'Hara}},\ and\ \citenamefont
  {Smith}}]{Mason_1972_JPB.5.169}%
  \BibitemOpen
  \bibfield  {author} {\bibinfo {author} {\bibfnamefont {E.~A.}\ \bibnamefont
  {Mason}}, \bibinfo {author} {\bibfnamefont {H.}~\bibnamefont {{O'Hara}}},\
  and\ \bibinfo {author} {\bibfnamefont {F.~J.}\ \bibnamefont {Smith}},\
  }\href@noop {} {\bibfield  {journal} {\bibinfo  {journal} {J. Phys. B: At.
  Mol. Phys.}\ }\textbf {\bibinfo {volume} {5}},\ \bibinfo {pages} {169}
  (\bibinfo {year} {1972})}\BibitemShut {NoStop}%
\bibitem [{\citenamefont {Nelson}\ \emph {et~al.}(2001)\citenamefont {Nelson},
  \citenamefont {Benhenni}, \citenamefont {Yousfi},\ and\ \citenamefont
  {Eichwald}}]{Nelson_2001_JPD.34.3247}%
  \BibitemOpen
  \bibfield  {author} {\bibinfo {author} {\bibfnamefont {D.}~\bibnamefont
  {Nelson}}, \bibinfo {author} {\bibfnamefont {M.}~\bibnamefont {Benhenni}},
  \bibinfo {author} {\bibfnamefont {M.}~\bibnamefont {Yousfi}},\ and\ \bibinfo
  {author} {\bibfnamefont {O.}~\bibnamefont {Eichwald}},\ }\href@noop {}
  {\bibfield  {journal} {\bibinfo  {journal} {J. Phys. D: Appl. Phys.}\
  }\textbf {\bibinfo {volume} {34}},\ \bibinfo {pages} {3247} (\bibinfo {year}
  {2001})}\BibitemShut {NoStop}%
\bibitem [{\citenamefont {Catlow}\ \emph {et~al.}(1982)\citenamefont {Catlow},
  \citenamefont {Dixon},\ and\ \citenamefont
  {Mackrodt}}]{Catlow_1982_Interionic}%
  \BibitemOpen
  \bibfield  {author} {\bibinfo {author} {\bibfnamefont {C.~R.~A.}\
  \bibnamefont {Catlow}}, \bibinfo {author} {\bibfnamefont {M.}~\bibnamefont
  {Dixon}},\ and\ \bibinfo {author} {\bibfnamefont {W.~C.}\ \bibnamefont
  {Mackrodt}},\ }in\ \href@noop {} {\emph {\bibinfo {booktitle} {Computer
  Simulation of Solids. Lecture Notes in Physics, vol. 166}}},\ \bibinfo
  {editor} {edited by\ \bibinfo {editor} {\bibfnamefont {C.~R.~A.}\
  \bibnamefont {Catlow}}\ and\ \bibinfo {editor} {\bibfnamefont {W.~C.}\
  \bibnamefont {Mackrodt}}}\ (\bibinfo  {publisher} {Springer, Berlin,
  Heidelberg},\ \bibinfo {year} {1982})\ pp.\ \bibinfo {pages}
  {130--161}\BibitemShut {NoStop}%
\bibitem [{\citenamefont {Gillan}(1986)}]{Gillan_1986_JPC.19.3391}%
  \BibitemOpen
  \bibfield  {author} {\bibinfo {author} {\bibfnamefont {M.~J.}\ \bibnamefont
  {Gillan}},\ }\href@noop {} {\bibfield  {journal} {\bibinfo  {journal} {J.
  Phys. C: Solid State Phys.}\ }\textbf {\bibinfo {volume} {19}},\ \bibinfo
  {pages} {3391} (\bibinfo {year} {1986})}\BibitemShut {NoStop}%
\bibitem [{\citenamefont {{van Beest}}\ \emph {et~al.}(1990)\citenamefont {{van
  Beest}}, \citenamefont {Kramer},\ and\ \citenamefont {{van
  Santen}}}]{BKS_SiO2_1990_PRL}%
  \BibitemOpen
  \bibfield  {author} {\bibinfo {author} {\bibfnamefont {B.~W.~H.}\
  \bibnamefont {{van Beest}}}, \bibinfo {author} {\bibfnamefont {G.~J.}\
  \bibnamefont {Kramer}},\ and\ \bibinfo {author} {\bibfnamefont {R.~A.}\
  \bibnamefont {{van Santen}}},\ }\href@noop {} {\bibfield  {journal} {\bibinfo
   {journal} {Phys. Rev. Lett.}\ }\textbf {\bibinfo {volume} {64}},\ \bibinfo
  {pages} {1955} (\bibinfo {year} {1990})}\BibitemShut {NoStop}%
\bibitem [{\citenamefont {Sundararaman}\ \emph {et~al.}(2018)\citenamefont
  {Sundararaman}, \citenamefont {Huang}, \citenamefont {Ispas},\ and\
  \citenamefont {Kob}}]{Sundararaman_2018_JCP.148.194504}%
  \BibitemOpen
  \bibfield  {author} {\bibinfo {author} {\bibfnamefont {S.}~\bibnamefont
  {Sundararaman}}, \bibinfo {author} {\bibfnamefont {L.}~\bibnamefont {Huang}},
  \bibinfo {author} {\bibfnamefont {S.}~\bibnamefont {Ispas}},\ and\ \bibinfo
  {author} {\bibfnamefont {W.}~\bibnamefont {Kob}},\ }\href@noop {} {\bibfield
  {journal} {\bibinfo  {journal} {J. Chem. Phys.}\ }\textbf {\bibinfo {volume}
  {148}},\ \bibinfo {pages} {194504} (\bibinfo {year} {2018})}\BibitemShut
  {NoStop}%
\bibitem [{\citenamefont {Ewald}(1921)}]{Ewald_1921_AnnPhys.369.253}%
  \BibitemOpen
  \bibfield  {author} {\bibinfo {author} {\bibfnamefont {P.}~\bibnamefont
  {Ewald}},\ }\href@noop {} {\bibfield  {journal} {\bibinfo  {journal} {Ann.
  Phys.}\ }\textbf {\bibinfo {volume} {369}},\ \bibinfo {pages} {253} (\bibinfo
  {year} {1921})}\BibitemShut {NoStop}%
\bibitem [{\citenamefont {Solov'yov}\ \emph
  {et~al.}(2017{\natexlab{b}})\citenamefont {Solov'yov}, \citenamefont
  {Sushko},\ and\ \citenamefont {Solov'yov}}]{MBNExplorer_UserGuide}%
  \BibitemOpen
  \bibfield  {author} {\bibinfo {author} {\bibfnamefont {I.~A.}\ \bibnamefont
  {Solov'yov}}, \bibinfo {author} {\bibfnamefont {G.}~\bibnamefont {Sushko}},\
  and\ \bibinfo {author} {\bibfnamefont {A.~V.}\ \bibnamefont {Solov'yov}},\
  }\href@noop {} {\emph {\bibinfo {title} {{MBN Explorer User's Guide (Version
  3.0)}}}}\ (\bibinfo  {publisher} {MesoBioNano Science Publishing},\ \bibinfo
  {address} {Frankfurt am Main},\ \bibinfo {year} {2017})\BibitemShut {NoStop}%
\bibitem [{\citenamefont {Liu}\ \emph {et~al.}(2020)\citenamefont {Liu},
  \citenamefont {Li}, \citenamefont {Fu}, \citenamefont {Li},\ and\
  \citenamefont {Bauchy}}]{Liu_2020_JCP.152.051101}%
  \BibitemOpen
  \bibfield  {author} {\bibinfo {author} {\bibfnamefont {H.}~\bibnamefont
  {Liu}}, \bibinfo {author} {\bibfnamefont {Y.}~\bibnamefont {Li}}, \bibinfo
  {author} {\bibfnamefont {Z.}~\bibnamefont {Fu}}, \bibinfo {author}
  {\bibfnamefont {K.}~\bibnamefont {Li}},\ and\ \bibinfo {author}
  {\bibfnamefont {M.}~\bibnamefont {Bauchy}},\ }\href@noop {} {\bibfield
  {journal} {\bibinfo  {journal} {J. Chem. Phys.}\ }\textbf {\bibinfo {volume}
  {152}},\ \bibinfo {pages} {051101} (\bibinfo {year} {2020})}\BibitemShut
  {NoStop}%
\bibitem [{\citenamefont {Woodcock}\ \emph {et~al.}(1976)\citenamefont
  {Woodcock}, \citenamefont {Angell},\ and\ \citenamefont
  {Cheeseman}}]{Woodcock_1976_JCP.65.1565}%
  \BibitemOpen
  \bibfield  {author} {\bibinfo {author} {\bibfnamefont {L.~V.}\ \bibnamefont
  {Woodcock}}, \bibinfo {author} {\bibfnamefont {C.~A.}\ \bibnamefont
  {Angell}},\ and\ \bibinfo {author} {\bibfnamefont {P.}~\bibnamefont
  {Cheeseman}},\ }\href@noop {} {\bibfield  {journal} {\bibinfo  {journal} {J.
  Chem. Phys.}\ }\textbf {\bibinfo {volume} {65}},\ \bibinfo {pages} {1565}
  (\bibinfo {year} {1976})}\BibitemShut {NoStop}%
\bibitem [{\citenamefont {Kramer}\ \emph {et~al.}(1991)\citenamefont {Kramer},
  \citenamefont {Farragher}, \citenamefont {{van Beest}},\ and\ \citenamefont
  {{van Santen}}}]{Kramer_1991_PRB.43.5068}%
  \BibitemOpen
  \bibfield  {author} {\bibinfo {author} {\bibfnamefont {G.~J.}\ \bibnamefont
  {Kramer}}, \bibinfo {author} {\bibfnamefont {N.~P.}\ \bibnamefont
  {Farragher}}, \bibinfo {author} {\bibfnamefont {B.~W.~H.}\ \bibnamefont {{van
  Beest}}},\ and\ \bibinfo {author} {\bibfnamefont {R.~A.}\ \bibnamefont {{van
  Santen}}},\ }\href@noop {} {\bibfield  {journal} {\bibinfo  {journal} {Phys.
  Rev. B}\ }\textbf {\bibinfo {volume} {43}},\ \bibinfo {pages} {5068}
  (\bibinfo {year} {1991})}\BibitemShut {NoStop}%
\bibitem [{\citenamefont {Alnemrat}\ and\ \citenamefont
  {Tomlinson}(2023)}]{Alnemrat_2023}%
  \BibitemOpen
  \bibfield  {author} {\bibinfo {author} {\bibfnamefont {S.}~\bibnamefont
  {Alnemrat}}\ and\ \bibinfo {author} {\bibfnamefont {W.~W.}\ \bibnamefont
  {Tomlinson}},\ }\href@noop {} {\bibfield  {journal} {\bibinfo  {journal} {J.
  Magn. Magn. Mater.}\ }\textbf {\bibinfo {volume} {566}},\ \bibinfo {pages}
  {170317} (\bibinfo {year} {2023})}\BibitemShut {NoStop}%
\bibitem [{\citenamefont {Verkhovtsev}\ \emph {et~al.}(2013)\citenamefont
  {Verkhovtsev}, \citenamefont {Hanauske}, \citenamefont {Yakubovich},\ and\
  \citenamefont {Solov'yov}}]{Verkhovtsev_2013_TiNi_clusters}%
  \BibitemOpen
  \bibfield  {author} {\bibinfo {author} {\bibfnamefont {A.~V.}\ \bibnamefont
  {Verkhovtsev}}, \bibinfo {author} {\bibfnamefont {M.}~\bibnamefont
  {Hanauske}}, \bibinfo {author} {\bibfnamefont {A.~V.}\ \bibnamefont
  {Yakubovich}},\ and\ \bibinfo {author} {\bibfnamefont {A.~V.}\ \bibnamefont
  {Solov'yov}},\ }\href@noop {} {\bibfield  {journal} {\bibinfo  {journal}
  {Comput. Mater. Sci.}\ }\textbf {\bibinfo {volume} {76}},\ \bibinfo {pages}
  {80} (\bibinfo {year} {2013})}\BibitemShut {NoStop}%
\bibitem [{\citenamefont {Kochaev}(2017)}]{Kochaev2017-rm}%
  \BibitemOpen
  \bibfield  {author} {\bibinfo {author} {\bibfnamefont {A.}~\bibnamefont
  {Kochaev}},\ }\href@noop {} {\bibfield  {journal} {\bibinfo  {journal} {Phys.
  Rev. B.}\ }\textbf {\bibinfo {volume} {96}},\ \bibinfo {pages} {155428}
  (\bibinfo {year} {2017})}\BibitemShut {NoStop}%
\bibitem [{\citenamefont {Liu}\ and\ \citenamefont
  {Wu}(2015)}]{Liu_2015_JMaterRes.31.832}%
  \BibitemOpen
  \bibfield  {author} {\bibinfo {author} {\bibfnamefont {K.}~\bibnamefont
  {Liu}}\ and\ \bibinfo {author} {\bibfnamefont {J.}~\bibnamefont {Wu}},\
  }\href@noop {} {\bibfield  {journal} {\bibinfo  {journal} {J. Mater. Res.}\
  }\textbf {\bibinfo {volume} {31}},\ \bibinfo {pages} {832} (\bibinfo {year}
  {2015})}\BibitemShut {NoStop}%
\bibitem [{\citenamefont {Zhang}\ \emph {et~al.}(2022)\citenamefont {Zhang},
  \citenamefont {Zhang}, \citenamefont {Yang}, \citenamefont {Zhao},
  \citenamefont {Chen}, \citenamefont {Cheng}, \citenamefont {Rasheed},
  \citenamefont {Ma},\ and\ \citenamefont {Zhang}}]{Zhang_2022_JPCC.126.1094}%
  \BibitemOpen
  \bibfield  {author} {\bibinfo {author} {\bibfnamefont {B.}~\bibnamefont
  {Zhang}}, \bibinfo {author} {\bibfnamefont {L.}~\bibnamefont {Zhang}},
  \bibinfo {author} {\bibfnamefont {N.}~\bibnamefont {Yang}}, \bibinfo {author}
  {\bibfnamefont {X.}~\bibnamefont {Zhao}}, \bibinfo {author} {\bibfnamefont
  {C.}~\bibnamefont {Chen}}, \bibinfo {author} {\bibfnamefont {Y.}~\bibnamefont
  {Cheng}}, \bibinfo {author} {\bibfnamefont {I.}~\bibnamefont {Rasheed}},
  \bibinfo {author} {\bibfnamefont {L.}~\bibnamefont {Ma}},\ and\ \bibinfo
  {author} {\bibfnamefont {J.}~\bibnamefont {Zhang}},\ }\href@noop {}
  {\bibfield  {journal} {\bibinfo  {journal} {J. Phys. Chem. C}\ }\textbf
  {\bibinfo {volume} {126}},\ \bibinfo {pages} {1094} (\bibinfo {year}
  {2022})}\BibitemShut {NoStop}%
\bibitem [{\citenamefont {Lee}\ \emph {et~al.}(2008)\citenamefont {Lee},
  \citenamefont {Wei}, \citenamefont {Kysar},\ and\ \citenamefont
  {Hone}}]{Lee_2008_Science.321.385}%
  \BibitemOpen
  \bibfield  {author} {\bibinfo {author} {\bibfnamefont {C.}~\bibnamefont
  {Lee}}, \bibinfo {author} {\bibfnamefont {X.}~\bibnamefont {Wei}}, \bibinfo
  {author} {\bibfnamefont {J.~W.}\ \bibnamefont {Kysar}},\ and\ \bibinfo
  {author} {\bibfnamefont {J.}~\bibnamefont {Hone}},\ }\href@noop {} {\bibfield
   {journal} {\bibinfo  {journal} {Science}\ }\textbf {\bibinfo {volume}
  {321}},\ \bibinfo {pages} {385} (\bibinfo {year} {2008})}\BibitemShut
  {NoStop}%
\bibitem [{\citenamefont {Liu}\ \emph {et~al.}(2014)\citenamefont {Liu},
  \citenamefont {Yan}, \citenamefont {Chen}, \citenamefont {Fan}, \citenamefont
  {Sun}, \citenamefont {Suh}, \citenamefont {Fu}, \citenamefont {Lee},
  \citenamefont {Zhou}, \citenamefont {Tongay}, \citenamefont {Ji},
  \citenamefont {Neaton},\ and\ \citenamefont {Wu}}]{Liu2014-ub}%
  \BibitemOpen
  \bibfield  {author} {\bibinfo {author} {\bibfnamefont {K.}~\bibnamefont
  {Liu}}, \bibinfo {author} {\bibfnamefont {Q.}~\bibnamefont {Yan}}, \bibinfo
  {author} {\bibfnamefont {M.}~\bibnamefont {Chen}}, \bibinfo {author}
  {\bibfnamefont {W.}~\bibnamefont {Fan}}, \bibinfo {author} {\bibfnamefont
  {Y.}~\bibnamefont {Sun}}, \bibinfo {author} {\bibfnamefont {J.}~\bibnamefont
  {Suh}}, \bibinfo {author} {\bibfnamefont {D.}~\bibnamefont {Fu}}, \bibinfo
  {author} {\bibfnamefont {S.}~\bibnamefont {Lee}}, \bibinfo {author}
  {\bibfnamefont {J.}~\bibnamefont {Zhou}}, \bibinfo {author} {\bibfnamefont
  {S.}~\bibnamefont {Tongay}}, \bibinfo {author} {\bibfnamefont
  {J.}~\bibnamefont {Ji}}, \bibinfo {author} {\bibfnamefont {J.~B.}\
  \bibnamefont {Neaton}},\ and\ \bibinfo {author} {\bibfnamefont
  {J.}~\bibnamefont {Wu}},\ }\href@noop {} {\bibfield  {journal} {\bibinfo
  {journal} {Nano Lett.}\ }\textbf {\bibinfo {volume} {14}},\ \bibinfo {pages}
  {5097} (\bibinfo {year} {2014})}\BibitemShut {NoStop}%
\bibitem [{\citenamefont {Lide}(2003)}]{CRC_Handbook_Chemistry}%
  \BibitemOpen
  \bibinfo {editor} {\bibfnamefont {D.~S.}\ \bibnamefont {Lide}},\ ed.,\
  \href@noop {} {\emph {\bibinfo {title} {{CRC Handbook of Chemistry and
  Physics, 84th ed.}}}}\ (\bibinfo  {publisher} {CRC Press},\ \bibinfo
  {address} {Boca Raton},\ \bibinfo {year} {2003})\BibitemShut {NoStop}%
\bibitem [{\citenamefont {Luo}\ \emph {et~al.}(2019)\citenamefont {Luo},
  \citenamefont {Yao}, \citenamefont {Tian}, \citenamefont {Song},
  \citenamefont {Wang}, \citenamefont {Li}, \citenamefont {Xi}, \citenamefont
  {Li}, \citenamefont {Song},\ and\ \citenamefont
  {Li}}]{BingchengLuo_PNAS_2019}%
  \BibitemOpen
  \bibfield  {author} {\bibinfo {author} {\bibfnamefont {B.}~\bibnamefont
  {Luo}}, \bibinfo {author} {\bibfnamefont {Y.}~\bibnamefont {Yao}}, \bibinfo
  {author} {\bibfnamefont {E.}~\bibnamefont {Tian}}, \bibinfo {author}
  {\bibfnamefont {H.}~\bibnamefont {Song}}, \bibinfo {author} {\bibfnamefont
  {X.}~\bibnamefont {Wang}}, \bibinfo {author} {\bibfnamefont {G.}~\bibnamefont
  {Li}}, \bibinfo {author} {\bibfnamefont {K.}~\bibnamefont {Xi}}, \bibinfo
  {author} {\bibfnamefont {B.}~\bibnamefont {Li}}, \bibinfo {author}
  {\bibfnamefont {H.}~\bibnamefont {Song}},\ and\ \bibinfo {author}
  {\bibfnamefont {L.}~\bibnamefont {Li}},\ }\href@noop {} {\bibfield  {journal}
  {\bibinfo  {journal} {Proc. Natl. Acad. Sci. U.S.A.}\ }\textbf {\bibinfo
  {volume} {116}},\ \bibinfo {pages} {17213} (\bibinfo {year}
  {2019})}\BibitemShut {NoStop}%
\end{thebibliography}%





\end{document}